\documentclass[twocolumn,aps,prd,10pt]{revtex4-1}

\usepackage{amsmath}
\usepackage{amssymb}
\usepackage{amsfonts}
\usepackage{graphicx}

\newcommand{\gap}{\rule{0pt}{17pt}}
\newcommand{\etal}{{\it et~al}.}
\newcommand{\CC}{{\mathcal C}}
\newcommand{\DA}{\mathcal{A}}
\newcommand{\DB}{\mathcal{B}}

\newcommand{\EE}{\mathcal{E}}
\newcommand{\PP}{\mathcal{P}}

\newcommand{\MM}{\text{M}}
\newcommand{\GG}{\text{G}}
\newcommand{\II}{\text{I}}
\newcommand{\LS}{\text{LS}}
\newcommand{\sys}{\text{(sys)}}
\newcommand{\env}{\text{(env)}}
\newcommand{\TT}{\text{TT}}
\newcommand{\TOT}{\text{T}}
\newcommand{\dd}{\text{d}}
\newcommand{\ppp}{\\[3pt]}
\newcommand{\nppp}{\nonumber\\[3pt]}
\newcommand{\np}{\nonumber\\}
\newcommand{\rr}{\boldsymbol{r}}
\newcommand{\xx}{\boldsymbol{r}}
\newcommand{\av}[1]{\langle#1\rangle}
\newcommand{\tr}{\text{Tr}}
\newcommand{\mm}{\boldsymbol{m}}
\newcommand{\nn}{\boldsymbol{n}}
\newcommand{\NN}{\boldsymbol{N}}
\newcommand{\kk}{\boldsymbol{k}}
\newcommand{\LL}{{\mathcal L}}
\newcommand{\HH}{{\mathcal H}}
\newcommand{\hc}{\text{H.c.}}
\newcommand{\bra}[1]{\langle#1|}

\newcommand{\ket}[1]{|#1\rangle}
\newcommand{\Idash}{I\hspace{-4.6pt}\raisebox{1.5pt}{-\hspace{-2pt}-}}
\newcommand{\Sec}[1]{Sec.~\ref{#1}}
\newcommand{\Secs}[2]{Secs.~\ref{#1} and \ref{#2}}
\newcommand{\apx}[1]{Appendix~\ref{#1}}

\newcommand{\Ref}[1]{Ref.~\cite{#1}}
\newcommand{\Refs}[1]{Refs.~\cite{#1}}
\newcommand{\refs}[1]{~\cite{#1}}
\newcommand{\eq}[1]{~\eqref{#1}}

\newcommand{\Eq}[1]{Eq.~\eqref{#1}}
\newcommand{\Eqq}[2]{Eqs.~\eqref{#1} and \eqref{#2}}

\newcommand{\Eqs}[1]{Eqs.~\eqref{#1}}

\begin{document}

\title{Quantum coherence, radiance, and resistance of gravitational systems}


\author{Teodora Oniga}
\email{t.oniga@abdn.ac.uk}

\author{Charles H.-T. Wang}
\email{Corresponding author. c.wang@abdn.ac.uk}
\affiliation{Department of Physics,
University of Aberdeen, King's College, Aberdeen AB24 3UE, United Kingdom
\vspace{10 pt}}



\begin{abstract}
We develop a general framework for the open dynamics of an ensemble of quantum particles subject to spacetime fluctuations about the flat background. An arbitrary number of interacting bosonic and fermionic particles are considered. A systematic approach to the generation of gravitational waves in the quantum domain is presented that recovers known classical limits in terms of the quadrupole radiation formula and backreaction dissipation. Classical gravitational emission and absorption relations are quantized into their quantum field theoretical counterparts in terms of the corresponding operators and quantum ensemble averages. Certain arising consistency issues related to factor ordering have been addressed and resolved. Using the theoretical formulation established here with numerical simulations in the quantum regime, we discuss potential new effects including decoherence through the spontaneous emission of gravitons and collectively amplified radiation of gravitational waves by correlated quantum particles.
\\ \vspace{30 pt}
\end{abstract}


\maketitle

\section{Introduction}

Recent detection of gravitational waves~\cite{LIGO2016} has confirmed one of the most important predictions of general relativity. Their discovery is not only realizing the long-awaited gravitational wave astronomy~\cite{Schutz1999, Sesana2016} but also puts the quest for deeper and wider progress of fundamental physics in a new perspective~\cite{gauge2009}.
Despite their prevailing classical descriptions, the energy density of the observed gravitational waves, close to the source GW150914, is thought to be a small fraction of the Planck density~\cite{LIGO2016}. This suggests the effects of quantum gravity and Planck scale physics on gravitational waves are of interest for further investigations. Indeed, one asks: What can be learned about quantum gravity from gravitational waves?

Gravitons are quantized gravitational waves~\cite{AshtekarLoops1991} and carry the true dynamics of gravitational fields~\cite{Wang2005a, Wang2005b, Wang2006a}. Like photons, under vacuum fluctuations spontaneous emission of gravitons by energized quantum states undergoing decay and decoherence has also been postulated. In particular, substantial spontaneous emissions of gravitons in the early universe following inflation by the matter content subject to quantum-to-classical transitions may be responsible for entropy production, thermodynamic arrow of time, structure formation, and the emergence of the classical world~\cite{kiefer2000, Joos2003, Schlosshauer2008, kiefer2012, Lim2015}. The precise physical mechanisms involved in this chain of processes are however not fully understood at present.
The ongoing efforts to observe gravitons of cosmological origin as part of primordial and stochastic gravitational waves \cite{LIGO2009, LIGO2016b} are expected to provide evidence for the above scenarios having considerable implications on the interplay between cosmology, quantum gravity, and potentially the ultimate unified theory at the Planck scale.

Driven by the above significant developments with the need for increased conceptual understanding and technical tools, we report in this paper on a unified framework based on recent theoretical progress of generic gravitational decoherence~\cite{Oniga2016a, Oniga2016b, Oniga2016c}, and provide an application example using a confined gravitating many-particle system ready to be generalized. The theory and methodology are aimed at addressing a wide range of complex and collective quantum dynamical behaviours of realistic matter systems that may be isolated in space but open to spacetime fluctuations~(Sec.~\ref{sec:var}). A broad class of phenomena may be relevant, covering gravitational decoherence, radiation with reaction and dissipation, and their classical reductions. We show that the classical dynamical structure for gravitational radiation is largely preserved as the deterministic part of the quantum structure, that also acquires an additional quantum stochastic influence from the universal fluctuations of spacetime~(Sec.~\ref{sec:rad}).
The generation of gravitational waves in the quantum domain under our systematic approach based on the modern formalism of open quantum systems \cite{Breuer2002} is shown to recover classical limits.
In treating the quantum mechanisms for gravitational emission and absorption in terms of quantized operators and quantum ensemble averages, we have encountered certain factor ordering ambiguities, which have fortunately been resolved through consistency considerations~(Sec.~\ref{sec:dec}).
The established theoretical formulation, illustrated with numerical simulations, allows us to demonstrate novel gravitational radiative phenomena including the collectively amplified spontaneous emission of gravitons by a highly coherent state of identical bosonic particles, in close analogy with the superradiance of photons ~\cite{Dicke1954}~(Sec.~\ref{sec:col}). Towards the end, we conclude this work with a summary of its results, implications, and future prospects~(Sec.~\ref{sec:con}).

In this work, we will consider the lowest order quantum gravitational effects consistent with the effective quantum field theory approach to general relativity\refs{Burgess2004}. At low energy, much less than the Planck scale, this description allows one to analyze the propagations of gravitons with matter interactions using linearized quantum gravity to be adopted below, without concerning the nonrenormalizability of gravity\refs{Arteaga2004}. Although such a restricted framework does not capture higher order quantum gravity effects, it is a significant necessary step in making progress towards a full quantum gravitational description, which has been useful in probing low-energy quantum gravitational decoherence\refs{Oniga2016a, Blencowe2013}. A better understanding of the physical effects of linearized quantum gravity may also guide the connections between a fuller theory of e.g. loop quantum gravity\refs{Ashtekar2017} with the real world. It is also sufficient to prove the quantum nature of gravity using linearized quantum gravity on the more accessible laboratory scales. Therefore, further theoretical and experimental understandings of linearized quantum gravity effects may bear important implications for full quantum gravity.

Additionally, the linearized quantum gravity framework serves as a tradeoff to suspend the problem of time in quantum gravity with full general covariance\refs{Isham1992}, by providing a background Minkowski metric $\eta_{\mu\nu}=$ diag$(-1,1,1,1)$ with Lorentz coordinates $(x^\mu)=(t,x,y,z)$, using Greek indices $\mu,\nu,\ldots=0,1,2,3$. When the metric is perturbed by a weak compact gravitational system and weak gravitational waves, these coordinates behave as mean asymptotic Lorentz coordinates for an observer distantly  exterior to the system. Such time $t=x^0$ may be measured e.g. by a laboratory which is stationary ``relative to a remote star.'' This way, while making no claims to resolve the ambiguity of time measurement often encountered in the context of quantum state reduction models\refs{Hu2013, Bassi2017}, prominently by Penrose\refs{Penrose1996}, we circumvent similar discussions with the above choice of time. Its physical consistency and usefulness within the linearized gravity approximation can be justified by the recovery of the classical limits of the quadrupole radiation formula and backreaction dissipation for gravitational waves from our quantum derivations, as required by the correspondence principle. See \Secs{sec:rad}{sec:dec}.

In what follows, apart from stated exceptions, we choose the relativistic units where the speed of light equals one, ${c=1}$. We retain in particular the reduced Planck $\hbar$ and Newtonian $G$ constants  to manifest quantum and gravitational couplings. Spatial coordinates in the Cartesian basis are indexed with Latin letters $i,j,\ldots=1,2,3$.
Summation over repeated indices is implied should no risk of confusion arise.
The time derivative, trace-reversion, Hermitian and complex conjugates are denoted by an over-dot $(\,\dot{}\,)$, over-bar $(\,\bar{}\,)$, superscripts $(^\dag)$ and  $(^*)$ respectively.
Symbols $H$ and $L$ are used for the Hamiltonian and Lagrangian with calligraphic type  $\HH$ and $\LL$ standing for their densities respectively.

\section{Covariant and canonical variables of matter-gravity systems}
\label{sec:var}


We start by considering the quantum dynamics of a (multicomponent) matter field $\varphi$ weakly coupled to gravity described by an action functional that can be approximated with
\begin{eqnarray}
S_\MM[\varphi,g_{\alpha\beta}]
&\approx&
S_\MM[\varphi,\eta_{\alpha\beta}]
+
\frac12
\int h_{\mu\nu} T^{\mu\nu}\,\dd^4 x
\label{Sapx}
\gap\end{eqnarray}
where the spacetime metric takes the perturbative form
$g_{\mu\nu}=\eta_{\mu\nu}+h_{\mu\nu}$
and
\begin{eqnarray}
T^{\mu\nu}
&=&
2\frac{\delta S_\MM}{\delta g_{\mu\nu}}
\Big|_{g=\eta}
\label{TS}
\gap\end{eqnarray}
is the stress-energy tensor of the matter on the Minkowski background.
Since the matter action $S_\MM[\varphi,g_{\alpha\beta}]$ above may depend on the derivatives of the metric, thereby accommodating spin connection for Dirac fields  \cite{Tucker1987, Tucker1995}, in this work we can extend the validity of the gravitational influence functional derived in \cite{Oniga2016a} for fermionic  as well as bosonic particles.

The expansion~\eqref{Sapx} gives rise to the matter Lagrangian of the form
\begin{eqnarray}
\LL_\MM
&=&
\LL^\sys_\MM(\varphi,\varphi_{,\alpha})
+
\LL_\II(\varphi,\varphi_{,\alpha},h_{\alpha\beta})
\gap\end{eqnarray}
where $\LL^\sys_\MM$, as the integrand of $S_\MM[\varphi,\eta_{\alpha\beta}]$,
describes the dynamics of the unperturbed matter system when gravity is switched off, and
\begin{eqnarray}
\LL_\II
&=&
\frac12 h_{\mu\nu}T^{\mu\nu}
\gap\end{eqnarray}
describes both the self interaction of matter through gravity, when switched on, as well as its gravitational interaction with the environment.
The total Lagrangian density
$
\LL_\TOT
=
\LL_\MM + \LL_\GG
$
in terms of $\LL_\GG=(16\pi G)^{-1}R$  yields the linearized Einstein equation
\begin{eqnarray}
G_{\mu\nu} = 8\pi G\,  T_{\mu\nu}
\label{eineqb}
\gap\end{eqnarray}
using the second order perturbation of the scalar curvature $R=R^{(2)}[h_{\alpha\beta}]$ and the first order perturbation of the Einstein tensor $G_{\mu\nu}=G_{\mu\nu}^{(1)}[h_{\alpha\beta}]$ whose expressions can be found in Ref.~\cite{MTW1973}.

\vspace{10pt}

Note that the Einstein equation\eq{eineqb} based on which the time evolution of the system density matrix to be developed is up to first order in metric perturbations. To obtain such first order field equations, the corresponding gravitational Lagrangian is therefore second order in metric perturbations as the fields. Accordingly we have consistently used the second order perturbation of the scalar curvature to enter into the gravitational Lagrangian for linearized gravity above.

\vspace{10pt}

The resulting classical theory is invariant under the gauge transformation
$h_{\mu\nu} \rightarrow h_{\mu\nu} + \xi_{\mu,\nu} + \xi_{\nu,\mu}$
induced from the coordinate transformation $x^\mu\to x^\mu-\xi^\mu$ for arbitrary displacement functions $\xi^\mu=(\xi,\xi^i)$.

\vspace{15pt}

To establish connection with the standard open system description in Hamiltonian formalism, where the perturbative interaction is assumed small, we introduce the conjugate momentum $\varpi$ of the matter field $\varphi$ with respect to $\LL_\MM^\sys$ and obtain the corresponding matter Hamiltonian density
\begin{eqnarray}\gap
\HH_\MM = \HH_\MM^\sys+\HH_\II
\label{MMI}
\gap\end{eqnarray}
where
$\HH_\MM^\sys=\varpi\dot{\varphi}-\LL_\MM^\sys$
and
\begin{eqnarray}
\HH_\II
&=&
-\frac12 h_{\mu\nu}T^{\mu\nu}.
\label{HI}
\gap\end{eqnarray}

\vspace{15pt}

The Hamiltonian density of linearized gravity
$\HH_\GG = p_{ij}h_{ij,0} - \LL_\GG$
takes the ADM form~\cite{ADM1962}
\begin{eqnarray}\gap
\HH_\GG
&=&
\HH_\GG^\env + n\CC_\GG + n_i\CC_\GG^i
\label{HGM}
\gap\end{eqnarray}
where $\HH_\GG^\env$ contains kinetic- and potential-like terms quadratic in $p^{ij}$ and $h_{ij}$ respectively counting for the positive energy of the environmental gravitational waves, and
\begin{eqnarray}
\CC_\GG
=
({16\pi G})^{-1}
\left( h_{ii,jj} - h_{ij,ij}\right),\;\;
\CC_\GG^i
=
-2p_{ij}{}_{,j}
\label{CCG}
\gap\end{eqnarray}
are first class constraints with Lagrangian multipliers
$n=-h_{00}/2$ and $n_i = h_{0i}$.
Therefore by using Eqs.~\eqref{MMI}, \eqref{HI} and \eqref{HGM}, the total Hamiltonian density $\HH=\HH_\MM+\HH_\GG$ can be expressed as
\begin{eqnarray}\gap
\HH_\TOT
&=&
\HH_\MM^\sys
+
\HH_\GG
+
\HH_\II
\label{HHGMI}
\ppp
&=&
\HH_\MM^\sys
+
\HH_\GG^\env
-
\frac12 h_{ij}T^{ij}
+
n\CC + n_i\CC^i
\label{HHGM}
\gap\end{eqnarray}
with the second line~\eqref{HHGM} above taking an overall ADM form using the constraints
\begin{eqnarray}\gap
\CC = \CC_\GG + \CC_\MM,\;\;
\CC^i = \CC_\GG^i + \CC_\MM^i
\label{CCGM}
\gap\end{eqnarray}
including the matter contribution

\newpage
\begin{eqnarray}\gap
\CC_\MM= T^{00},\;\;
\CC_\MM^i=-T^{0i}.
\label{CCM}
\gap\end{eqnarray}


This Hamiltonian formulation enables the gauge transformations of all dynamical variables of the matter-gravity system to be generated by the first class constraints $\CC$ and $\CC^i$ through canonical transformations.

\section{Radiation, reception, and reaction of gravitational waves}
\label{sec:rad}

In the Lorenz gauge $\bar{h}_{\mu\nu}{}^{,\nu}=0$, the linearized Einstein equation~ \eqref{eineqb} takes the form
\begin{eqnarray}
h_{\mu\nu,\alpha}{}^\alpha = - 16\pi G  \bar{T}_{\mu\nu} \,.
\label{lfeq3}
\gap\end{eqnarray}
with solutions naturally separated into
\begin{eqnarray}
h_{\mu\nu} = h^\sys_{\mu\nu} + h^\env_{\mu\nu}.
\label{bug}
\gap\end{eqnarray}
The first term above is an inhomogeneous solution combined from
\begin{eqnarray}
h^\sys_{\mu\nu}(\rr,t) =
4 G \int \dd^3 x'
\frac{\bar{T}_{\mu\nu}(\rr',t-\epsilon|\rr-\rr'|)}{|\rr-\rr'|}
\label{ret}
\gap\end{eqnarray}
using the spatial position vector $\rr$ with norm $|\rr|=r$,
for $\epsilon=1$ as a retarded potential, and $\epsilon=-1$ as an advanced potential, describing respectively the radiation and reception of gravitational waves by the mater system.


When the usual outgoing-wave boundary condition is applied with $\epsilon=1$, the amplitude $h^\sys_{\mu\nu}$ appears to ``leak into the environment'' and becomes observable gravitational waves, though technically $h^\sys_{\mu\nu}$ is tied to the matter system, and is not part of the environment.
Likewise, if the less familiar though physically possible ingoing-wave boundary condition is applied with $\epsilon=-1$, the amplitude $h^\sys_{\mu\nu}$ appears to be ``sucked from the environment'', though again $h^\sys_{\mu\nu}$ is technically not part of the environment.

The second term $h^\env_{\mu\nu}$ of Eq.~\eqref{bug} above satisfies the homogeneous part of Eq.~\eqref{lfeq3} and describes the environmental gravitational waves. As such, the addition transverse-traceless (TT) condition can be applied to
$h^\env_{\mu\nu}$. Since $h^\env_{ij}$ is independent of the mater system, it carries the dynamical degrees of freedom of gravity.

The orthogonality of the TT decomposition allows us to split the interacting Hamiltonian density \eqref{HI} into
\begin{eqnarray}
\HH_\II = \HH_\II^\sys + \HH_\II^\env
\label{IHW}
\gap\end{eqnarray}
where
\begin{eqnarray}
\HH_\II^\sys
=
-\frac12 h^\sys_{\mu\nu} T^{\mu\nu}
\label{HIU}
\gap\end{eqnarray}
describing the self-gravity of the matter system and
\begin{eqnarray}
\HH_\II^\env = -\frac12 h^\env_{ij} \tau_{ij}
\label{HIW}
\gap\end{eqnarray}
in terms of the TT stress tensor $\tau_{ij}=T^\TT_{ij}$, describing the coupling between the matter system and the environmental gravitational waves.

\vspace{10pt}

The interacting matter system
\begin{eqnarray}
\HH_\MM = \HH_\MM^\sys+\HH_\II^\sys
\label{MMII}
\gap\end{eqnarray}
obtained from Eq.~\eqref{MMI} by incorporating self-gravity Eq.~\eqref{HIU} and hence turning off the environmental gravity, i.e. $h^\env_{ij}=0$, provides a closed dynamics for the classical radiation (or reception) of gravitational waves whose wave amplitude is determined by Eq.~\eqref{ret}.
For a nonrelativistic compact matter system of size $r^\sys$ much less than the wavelength, one obtains the TT part of this wave amplitude to be
\begin{eqnarray}
h^\TT_{ij}(t)
&=&
\frac{2G}{r}\,\ddot{\Idash}^{\,\TT}_{ij}(t-\epsilon r)
\label{gwh}
\gap\end{eqnarray}
at a distance $r \gg r^\sys$ from the matter system having the reduced quadrupole moment
\begin{eqnarray}
\Idash_{ij}
&=&
\int\dd^3x\,\big(x^i x^j - \frac{1}{3}\,\delta_{ij}\, r^2\big)\,T^{00}(\rr,t).
\label{qij}
\gap\end{eqnarray}

The average radiation (or reception) power can be derived from integrating the total flux associated with Eq.~\eqref{gwh} using the gravitational wave energy density
\begin{eqnarray}
\EE
&=&
\frac1{32\pi G}\,\av{\,\dot{h}^\TT_{ij}\,\dot{h}^\TT_{ij}\,}
\label{GWE}
\gap\end{eqnarray}
to be the well-known quadrupole gravitational radiation formula
\begin{eqnarray}
\PP
&=&
\frac{G}{5}\,\av{\dddot{\Idash}_{ij}\dddot{\Idash}_{ij}}
\label{qhpwr}
\gap\end{eqnarray}
where $\av{\cdot}$ denotes classical averaging, which in principle applies for gravitational reception as well.

Since the above gravitationally interacting matter system is closed, deterministic and conservative, the gravitational wave energy escaping to (or feeding from) infinity must involve balancing (anti-)dissipation. This mechanism, at the classical level~\cite{Maggiore2008}, is indeed provided by the backreaction from the gravitational wave amplitude $h^\sys_{\mu\nu}$ through its time retardation (or advance) induced effective (anti-)damping using Eqs.~\eqref{ret} and \eqref{HIU}.

\section{Decoherence via spontaneous emission and absorption of gravitons}
\label{sec:dec}

The preceding paradigm for the radiation, reception and reaction of gravitational waves changes drastically when the fundamental quantum properties of matter and gravity are taken into account. The field theoretical nature of linearized gravity means that after quantization there is a permanent fluctuating gravitational background even at zero temperature. The ambient spacetime fluctuations couple universally to all matter systems through the environmental interaction term $\HH_\II^\env$ given by Eq.~\eqref{HIW}. Like $\HH_\II^\sys$ in Eq.~\eqref{HIU}, this term can drain energy e.g. at a low environmental temperature, as well as pump energy e.g. at a high environmental temperature. Therefore, for a quantized gravitating system, there are now two channels of energy flow from the system: radiation reaction with a deterministic character and spacetime fluctuations with a stochastic character and hence a capacity to decohere.
It may be physically conceivable that the exchange of gravitational energies, for classical-like macroscopic systems with fluctuations smoothed out, is dominated by radiation reaction, whereas for quantumlike microscopic systems with diminishing time retardation or advance inside the system, is dominated by spacetime fluctuations.


To quantize the total matter-gravity system while preserving gauge invariance, we carry out Dirac's canonical quantization of constrained system~\cite{Dirac1964} based on the Hamiltonian density~\eqref{HHGM} in the Heisenberg picture~\cite{Oniga2016a}, where the operator forms of the first class constraints $\CC$ and $\CC^i$ given by Eq.~\eqref{CCGM} become quantum generators of gauge transformation. Accordingly, physical states $\ket{\psi}$ are required to be gauge invariant by satisfying the quantum constraints
\begin{eqnarray}
\CC \ket{\psi}=0
\,,\;
\CC^i \ket{\psi}=0.
\label{qCCipsi}
\gap\end{eqnarray}
In what follows, our perturbative approach would naturally admit a ``Dirac-Fock'' description of quantization, for which it has been shown that only the positive frequency modes of the constraints are required to annihilate physical states. See e.g. \Ref{Antoniadis1997} for relevant discussions and further details on the consistent Dirac quantization using the Fock representations.

The canonical variable operators acting on physical states satisfying Eq.~\eqref{qCCipsi} then evolve in time according to the quantum Heisenberg equations, which are equivalent to the quantum linearized Einstein equation~\eqref{eineqb}. In this formalism, supplementary relations can be used to restrict gauge redundances, as the quantum form of gauge conditions at no expense of breaking gauge invariance as gauge transformations can still be generated by $\CC$ and $\CC^i$~\cite{Oniga2016a}.

In this sense, to establish the influence of the quantum gravitational environment on the matter system, it is useful to work in the quantum Lorenz gauge so that the metric perturbation operator $h_{\mu\nu}$ satisfy quantized Eq.~\eqref{lfeq3} with solutions also separated in the same manner as Eq.~\eqref{bug}.
Using quantized Eqs.~\eqref{HGM}, \eqref{HHGMI}, and Eq.~\eqref{IHW}, and considering only physical states satisfying the quantum constraints~\eqref{qCCipsi}, we obtain the total Hamiltonian that governs the evolution and coupling of the matter-gravity system as follows
\begin{eqnarray}
H_\TOT
&=&
H_\MM^\sys
+
H_\GG^\env
+
H_\II
\nppp
&=&
H_\MM^\sys
+
H_\GG^\env
+
H_\II^\sys
+
H_\II^\env.
\label{IHWII}
\gap\end{eqnarray}

To investigate the dynamics of matter-gravity coupling and the resulting radiation, decoherence and dissipation, we will from now on employ the interaction picture where the interaction Hamiltonian $H_\II$, consisting of self ($H_\II^\sys$) and environmental ($H_\II^\env$) gravity contributions, generates the time evolution of quantum states.
We consider the fluctuating spacetime to resemble an infinite reservoir in which environmental gravitons with frequencies $\omega$ are maintained in an equilibrium Gaussian state with a distribution function $N(\omega)$ described by a gravitational density matrix $\rho_\GG$. For thermal equilibrium $N(\omega)$ is given by the Planck distribution function and for the zero-point spacetime fluctuations $N(\omega)$ vanishes.

In terms of the total density matrix $\rho_\TOT(t)$ of the matter system and the gravitational environment, the total time evolution is determined by the Liouville-von Neumann equation
\begin{equation}
\dot\rho_\TOT
=
-\frac{i}{\hbar}\, [H_\II,\rho_\TOT].
\label{drho}
\gap\end{equation}
The density matrix describing the statistical state of the matter system is reduced from the total system by averaging over the ensembles of the gravitational reservoir through the partial trace
\begin{equation}
\rho_\MM = \tr_\GG(\rho_\TOT).
\label{rhoM}
\gap\end{equation}

For a matter system initially untangled with the gravitational environment at $t=0$, when the total state takes the factored form
\begin{equation}
\rho_\TOT(0) = \rho_\MM(0) \otimes \rho_\GG
\label{ini}
\gap\end{equation}
which may later develop entanglement with the environment,
its reduced dynamical evolution is generated by the non-Markovian master equation \begin{eqnarray}
\hspace{-30pt}
\dot\rho
&=&
-\frac{i}{\hbar} [H_\II^\sys, \rho]
-\frac{8\pi G}{\hbar}
\int\!\! \frac{\dd^3 k}{2(2\pi)^3k}
\nppp&&
\times\,
\Big \{
\int_{0}^{t} \!\!\dd t'
e^{-i k (t - t')}
\big(
[
\tau^\dag_{ij} (\kk, t),\,
\tau_{ij}(\kk, t') \rho
]
\nppp
&&
\,+\,N(\omega_k)\,
[
\tau^\dag_{ij} (\kk, t),\,
[
\tau_{ij}(\kk, t'),\,
\rho
]]
\big)
+\hc
\Big\}
\label{maseqn}
\gap\end{eqnarray}
using Eqs.~\eqref{drho}, \eqref{rhoM}, \eqref{ini}, and the gauge invariant gravitational influence functional techniques~\cite{Oniga2016a}. Above,
$\rho=\rho_\MM$ abbreviates the matter system density matrix, $\omega_k=k=|\kk|$ denotes the environmental graviton frequency associated with wave vector $\kk$, and
\begin{eqnarray}
\tau_{ij}(\kk, t)
&=&
\int\!
\tau_{ij}(\rr, t)\,
e^{-i \kk\cdot\rr}\,\dd^3x
\gap\end{eqnarray}
are operators Fourier-transformed from quantized $\tau_{ij}(\rr, t)$ introduced in Eq.~\eqref{HIW}, which have been normal-ordered with particle nonconservation terms neglected in the low energy domain being considered.

Notably, \Eq{maseqn} constitutes an integrodifferential equation satisfied by the Dyson series solutions of the spacetime-ensemble averaged \Eq{drho}, whose time-nonlocality gives rise to non-Markovianity\refs{Oniga2016a}. In accord with the perturbation theory of the non-Markovian dynamics of open quantum systems\refs{Fleming2012}, the order of coupling in such a series expansion increases consistently by one, for each Dyson expansion order, with an extra time integral.

Similarly, in the standard perturbative scattering theory with a linear order coupling in the sense of time-local field equations, the transition amplitudes can be obtained from the Dyson expansions containing time-nonlocal integrals with nonlinear coupling orders, physical constraints permitting. For instance, the validity of such expansions for a scattering system may be limited by whether pair productions or other high-energy effects are evident. For weak gravitational systems being considered, the size of the dynamical metric perturbations should ultimately remain much less than order one for \Eq{maseqn} to be valid.

It is also worth remarking that, in deriving \Eq{maseqn}, the averaging over the Gaussian environment with zero-mean fluctuating gravitational fields assimilates the Dyson expansion into a cumulant expansion that terminates at the second order, making the non-Markovian master equation\eq{maseqn} truncation-free\refs{Oniga2016a}.

The second coupling order with fluctuating linearized gravity has also emerged previously in calculating transition amplitudes under a gravitational bath\refs{Schafer1980, Schafer1981} using Feynman's path integral approach\refs{Feynman1963}. Indeed, second-order master equations have been a prevalent feature for models of stochastic quantum evolutions under weak gravitational fluctuations\refs{Hu2013, Bassi2017}.


Here we investigate new nontrivial dynamical consequences of this master equation in a more general physical context, covering in particular radiation through quantum decoherence and dissipation for particles in confined states as opposed to free particles studied in Ref.~\cite{Oniga2016b}. Kinematically, the finite spatial extension of such a system permits the definitions of outgoing and ingoing gravitational waves. Dynamically, the coupling between these waves and the time evolution of the system results in their emissions (or absorptions) through Eq.~\eqref{ret} in general and Eq.~\eqref{gwh} for nonrelativistic systems. On quantization, these equations~\eqref{ret} and \eqref{gwh} become operator equations.

\vspace{10pt}

The average gravitational wave energy density expression~\eqref{GWE} then acquires quantum meaning by interpreting ${h}^\TT_{ij}$ there to be operators and averaging to be over quantum ensembles so that given a variable $v$ we have
\begin{eqnarray}
\av{v}=\tr(\,v\rho\,)
\label{var}
\gap\end{eqnarray}
using the matter system density matrix~$\rho$~\cite{Breuer2002}.
The quantum radiation formula also takes the same form as Eq.~\eqref{qhpwr} through quantized Eq.~\eqref{gwh}.
However, factor ordering requires some care here as the energy density related term
$\dot{h}^\TT_{ij}\,\dot{h}^\TT_{ij}$ in Eq.~\eqref{GWE} is normal-ordered.
Accordingly, when the reduced quadrupole moment operator given by quantized Eq.~\eqref{qij} is expanded in frequency modes
\begin{eqnarray}
\Idash_{ij}(t) =
a_{ij}(\omega)\, e^{-i\omega t} + a_{ij}^\dag(\omega)\, e^{i\omega t}
\gap\end{eqnarray}
for some operators $a_{ij}(\omega)$ with positive frequencies $\omega$,
the normal-like ordering of these operators
\begin{eqnarray}
a_{ij}\, a_{kl}^\dag
\to
a_{kl}^\dag\, a_{ij}
\label{anij}
\gap\end{eqnarray}
should be implemented for consistency.

Factor ordering for the interaction Hamiltonian $H_\II$ given by Eq.~\eqref{IHW} bears some fundamental significance. For electromagnetic radiative problems, it is known that different factor ordering for the analogous interaction Hamiltonian leads to physically distinct mixes and separations of effects from vacuum fluctuations and radiation reaction \cite{Ackerhalt1973, DDC1982, DDC1984, Scully1988, Menezes2015}.
Here, the gravitational coupling is constructed from the quantized general action~\eqref{Sapx} assumed to be Hermitian for any metric perturbation operator $h_{\mu\nu}$. It follows that, the interaction Hamiltonian $H_\II$ separated from $H_\MM$ with arbitrary $h_{\mu\nu}$ factor is necessarily Hermitian. Now, from Eq.~\eqref{IHW}, the interaction Hamiltonian $H_\II$ is the sum of the  environmental part $H_\II^\sys$ in Eq.~\eqref{HIW}, which is Hermitian as $h^\env_{ij}$ and $\tau_{ij}$ commute, and the system part~\eqref{HIU}, which is not readily Hermitian as $h^\sys_{\mu\nu}$ is related to time delayed or advanced $T^{\mu\nu}$ through  Eq.~\eqref{ret} and so may not commute with $T^{\mu\nu}$. Nonetheless, to achieve the Hermiticity of $H_\II^\sys$ and hence of $H_\II$, with the correct classical limit, the factor ordering for Eq.~\eqref{IHW} can be resolved symmetrically as follows
\begin{eqnarray}
\HH_\II^\sys
=
-\frac14 h^\sys_{\mu\nu} T^{\mu\nu}
-\frac14 T^{\mu\nu} h^\sys_{\mu\nu}.
\label{HIUQ}
\gap\end{eqnarray}

Similarly symmetrized interaction Hamiltonian \cite{DDC1982, DDC1984} has been applied in resolving the aforementioned factor ordering ambiguity in a wide range of problems involving electromagnetic fluctuations and radiation reaction. A recent related discussion and review can be found in Ref.~\cite{Menezes2015}.

\section{Collective radiation by confined identical particles}
\label{sec:col}

The theoretical framework established above is applied in this section, as an illustrative example, to the quantum gravitational decoherence and radiation of
a real, i.e., neutral, scalar field $\phi$ with mass $m$ and the associated inverse reduced Compton wavelength $\mu = m/\hbar$, subject to an external nongravitational potential $\nu(\rr)$ described by the Lagrangian density
\begin{eqnarray}
\LL
&=&
-
\frac12\,g^{\alpha\beta}\phi_{,\alpha}\phi_{,\beta}
-\Big(\frac{1}{2}+\nu\Big)\mu^2
\phi^2.
\label{LL}
\gap\end{eqnarray}
We focus on the newly formulated spontaneous emission of gravitons by nonrelativistic particles through environmental decoherence at zero temperature and highlight previously undiscovered collective gravitational radiation, which we will refer to as ``superradiance of gravitational waves'' that mirrors its original electromagnetic description~\cite{Dicke1954}.
As noted in Sec.~\ref{sec:dec} and by analogy with standard treatments in quantum optical systems~\cite{Breuer2002}, we assume the radiation process to be primarily due to spacetime fluctuations using $H_\II^\env$ by neglecting radiation reaction from self-gravity using $H_\II^\env$. As a result, quantum dissipation alone is responsible for the radiative loss of energy, which we verify explicitly for one particle excited in one dimension.
To consider the nonrelativistic dynamics of the scalar field representing nearly Newtonian particles we assume the potential energy to be much less than the mass energy so that
$\nu \ll 1$.

In the presence of weak gravity, we have $\eta_{\mu\nu} \to \eta_{\mu\nu} + h_{\mu\nu}$ with the proper coordinates
\begin{eqnarray}
x^i
&\to&
x^i+\frac12h^\env_{ij} x^j + O(h^\sys_{jk} x^l)
\gap\end{eqnarray}
by using \eqref{bug}, with related considerations discussed in Ref.~\cite{Maggiore2008}.
The resulting fluctuating potential in the TT gauge for free gravitational waves is given by
\begin{eqnarray}
\nu(x^i)
&\to&
\nu(x^i) + \frac{1}{2}h^\env_{ij} x^i \nu_{,j} + O(h^\sys \nu).
\label{v2v}
\gap\end{eqnarray}
The second and third terms in Eq.~\eqref{v2v} above contribute respectively to $\HH^\env$ in Eq.~\eqref{HIW} and $\HH^\sys$ in Eq.~\eqref{HIU}.
The appearance of gravitational wave induced potential fluctuations have also been discussed in \Refs{Schafer1980, Schafer1981}. However, if the confinement of particles is limited by free masses then the corresponding boundaries fluctuate in the proper coordinates instead of the TT coordinates~\cite{Oniga2016c}.

The system part of Eq.~\eqref{LL} yields the unperturbed quantum field equation
\begin{eqnarray}
\ddot\phi
=
\nabla^2\phi
-
(1+2\nu)\mu^2\phi
\label{hseqa}
\gap\end{eqnarray}
having solutions of the form
\begin{eqnarray}
\phi
=
\Psi_n(\rr)\, e^{-i\omega_n t} + \hc
\label{hPhi}
\gap\end{eqnarray}
with some orthogonal operators $\Psi_n(\rr)$.
Hence Eq.~\eqref{hseqa} reduces formally to the time-independent Schr\"odinger equation
\begin{eqnarray}
-\frac{\hbar^2}{2m}\nabla^2\Psi_n + V\Psi_n
=
E_n \Psi_n
\label{hschreq}
\gap\end{eqnarray}
where $V=m\,\nu(\rr)$ and
\begin{eqnarray}
E_n
=
\frac{1}{2m}(\hbar^2\omega_n^2-m^2)
\label{wmE}
\gap\end{eqnarray}
represent the potential and eigen energies respectively.


As a concrete physical configuration, let us consider an isotropic harmonic potential with frequency $\omega$:
\begin{eqnarray}
V
=
\frac12\,m\omega^2 r^2.
\gap\end{eqnarray}
In this case, Eq.~\eqref{hPhi} becomes
\begin{eqnarray}
\phi
&=&
\sqrt{\frac{\hbar}{2\omega_{\nn}}}\,
\big(
a_{\nn} e^{-i\omega_{\nn} t}
+
a_{\nn}^\dag  e^{i\omega_{\nn} t}
\big)
\psi_{\nn}(\rr)
\label{3hphi}
\gap\end{eqnarray}
using the multiple indices $\nn=(n_1,n_2,n_3)$ for $n_1,n_2,n_3=0,1,2\dots$, and functions
\begin{eqnarray}
\psi_{\nn}(\rr)
&=&
\psi_{n_1}(x)\psi_{n_2}(y)\psi_{n_3}(z)
\gap\end{eqnarray}
with the harmonic oscillator wave functions $\psi_n(x)$ and
\begin{eqnarray}
\omega_{\nn}
&=&
\mu +
(n_1+n_2+n_3)\,\omega
\label{3nome}
\gap\end{eqnarray}
arising from  the nonrelativistic limit of Eq.~\eqref{wmE}, where the corresponding ladder operators $a_{\nn}$ and $a_{\nn}^\dag$ are annihilation and creation operators respectively.

\vspace{10pt}

In terms of the TT projector $P_{ijkl}$ \cite{Oniga2016a}, the TT part of the stress-energy tensor follows from Eqs. \eqref{TS}, \eqref{LL} and \eqref{v2v} to be
\begin{eqnarray}
\tau_{ij}(\rr, t)
&=&
P_{ijkl}
\big(
\phi_{,k}\phi_{,l}
-
\mu^2\omega^2 x^k x^l \phi^2
\big)
\gap\end{eqnarray}
where the second contribution proportional to $\phi^2$ arises from the second term in \Eq{v2v}, which is induced from metric fluctuations having no electromagnetic analogue as discussed in \Sec{sec:dec}.
From this, by normal-ordering and neglecting particle nonconservation terms relevant only for higher energy scales,
we then obtain
\begin{eqnarray}
\tau_{ij}(\kk, t)
&=&
F_{ij}(\nn,\nn',\kk)\,
a_{\nn'}^\dag a_{\nn} e^{-i(\omega_{\nn}-\omega_{\nn'}) t}
\label{3ht10a}
\gap\end{eqnarray}
where
\begin{eqnarray}
F_{ij}(\nn',\nn,\kk)
&=&
\frac{\hbar}{\mu}
P_{ijkl}(\kk)
\times
\nppp&&
\hspace{-35pt}
\int \!\dd^3 x\,
\big(
\psi_{\nn',k}
\psi_{\nn,l}
-
\mu^2\omega^2 x^k x^l
\psi_{\nn'}
\psi_{\nn}
\big)
\label{d3hFun}
\gap\end{eqnarray}
using nonrelativistic approximation with $n_{\max}\omega \ll \mu$ as the kinetic energy is much less than the rest mass energy and related long transmitted gravitational wave length condition compared to the spatial extension of occupied harmonic modes.


To derive the gravitational analogue of the  quantum optical master equation for the particle system from the general master equation~\eqref{maseqn}, we carry out the Markov approximation~\cite{Breuer2002} as follows. First, we substitute Eq.~\eqref{3ht10a} into an integral in Eq.~\eqref{maseqn} to get
\begin{eqnarray}
&&
\int_{0}^{t}\dd t'\,
\tau_{ij}(\kk,t')
e^{-i k (t-t')}
=
F_{ij}(\nn,\nn',\kk)\,a_{\nn'}^\dag a_{\nn}
\nppp&&\hspace{35pt}
\times\;
e^{-i(\omega_{\nn}-\omega_{\nn'}) t}
\int_{0}^{t}\dd s\,
e^{-i (k-\omega_{\nn}+\omega_{\nn'}) s}.
\label{3ht10d}
\gap\end{eqnarray}
The nonlocality of this expression in time represents the non-Markov memory effect, which tends to fade away under environmental dissipation.
We ``forget'' this memory by taking the limit
$\int_{0}^{t} \dd s
\to
\int_{0}^{\infty} \dd s$,
as it does not affect post-transient dynamics, and apply the Sokhotski-Plemelj theorem
\begin{equation}
\int_0^\infty \dd s\, e^{-i \epsilon s}
= \pi \delta(\epsilon) - i \,\mathrm{P} \frac{1}{\epsilon}
\label{thm}
\gap\end{equation}
to Eq.~\eqref{3ht10d}, where $\mathrm{P}$ denotes the Cauchy principal value that gives rise to a nondissipative Hamiltonian $H^\env_\LS$ for the environmentally induced Lamb and Stark shifts of energy.
By analogy with quantum optics~\cite{Breuer2002}, we capture the leading radiative mechanisms by adopting the rotating wave approximation, neglecting self-gravity $H^\sys_\II$ and Lamb and Stark shift $H^\env_\LS$ Hamiltonians, when substituting the resulting Eq.~\eqref{3ht10d} back into Eq.~\eqref{maseqn}.

\vspace{10pt}

Although our general description covers both emission and absorption of gravitons, for a typical environment with a very low level of gravitational wave background, let us focus on the emission of gravitons in the following, leaving the absorption to a separate discussion\refs{Quinones2017}.
Thus we suppress the absorption of gravitons by setting their environmental distribution function ${N(\omega)=0}$, hence retaining merely zero-point fluctuations in the gravitational environment. The above considerations lead us to the gravitational quantum optical master equation
\begin{eqnarray}
\dot\rho
=
\frac{\Gamma}{2}
\big(
3\delta_{ik}\delta_{jl}
-
\delta_{ij}\delta_{kl}
\big)
\big(
A_{ij}
\rho
A^\dag_{kl}
-
\frac12\{
A^\dag_{ij}A_{kl},
\rho
\}
\big)
\np
\label{maseq}
\gap\end{eqnarray}
of the Lindblad form, with the transition rate coefficient
\begin{eqnarray}
\Gamma
&=&
\frac{32\,G\hbar\,\omega^3}{15\, c^5}
\label{Gamma0}
\gap\end{eqnarray}
where the speed of light $c$ has been reinstated,
and the associated Lindblad operators
\begin{eqnarray}
A_{ij}
&=&
\sum_{\nn}\sqrt{n_i (n_j -\delta_{ij})}\,
a^\dag_{\nn-\hat\nn_i-\hat\nn_j}a_{\nn}
\label{AAij}
\gap\end{eqnarray}
where $\hat\nn_1=(1,0,0), \hat\nn_2=(0,1,0), \hat\nn_3=(0,0,1)$.
It is important to note that although time $t$ in the original non-Markovian master equation~\eqref{maseqn} starts with an initially factored state~\eqref{ini} untangled with the environment, the Markov assumption used in arriving at Eq.~\eqref{maseq} has effectively pushed that initial time back to the infinite past whose memory is lost \cite{Breuer2002}, with time $t$ now reset to start from any new initial condition for the reduced matter state $\rho=\rho_\MM(t)$. The detailed derivation of \Eq{maseq} is given in \apx{app:maseq}.

The justification of the above Markov assumption necessarily requires the evolution time scale $\Delta t$ for master equation\eq{maseq} to be much greater than the system time scale $\tau=2\pi/\omega$, i.e. over many circles of the system oscillations, for \Eq{thm} to provide a good approximation to the last integral of \Eq{3ht10d}, where $k$ is fixed to be $2\omega$ by \Eq{2om} as shown in \apx{app:maseq}. Likewise, the rotating wave approximation requires $\Delta t \gg \tau$ for the evolution time scale $\Delta t$ to be long enough to average out oscillations on a faster time scale of $2\pi/\omega$. Therefore, for the system transition time scale using \Eq{Gamma0} to be validity we must require $1/\Gamma \gg \tau$ for a single particle system. This can be practically satisfied, thanks to the smallness of $\Gamma$ for conceivable oscillators. In a broad context of quantum Brownian motion\refs{Hu1992}, non-Markovianity can arise even without an integrodifferential structure and the justification of the Markov assumption may require more than time-scale comparisons. Nonetheless, for a large class of open quantum oscillator models, Markovian master equations are shown to often provide good approximations at sufficiently high temperature and for sufficiently weak system-environment coupling at low or zero temperature\refs{Breuer2002, Hu1992}. The latter condition amounts to $1/\Gamma \gg \tau$ stated above in our case. However, for collectively amplified transitions with a particle number $N$ to be discussed below, the condition beyond which non-Markovian effects could start to occur may become more stringent, as the transition rate scales with $N^2$.

Under the Markovian evolution using Eq.~\eqref{maseq} at zero temperature, an excited state $\rho$ decoheres and decays towards the ground state. In the process, gravitons are spontaneous emitted that carry the same amount of energy as being reduced from the matter system.
For example, let us take an arbitrary one-particle state $\rho$ with matrix elements
$\rho_{\nn,\nn'} = \bra{\nn}\rho\ket{\nn'}$ with $\ket{\nn'}$ as the state vector for the occupation of a harmonic mode $\nn$ by one particle. Then we obtain from Eqs.~\eqref{var} and \eqref{maseq} the dissipation power
\begin{eqnarray}
-\frac{\dd \av{H^\sys}}{\dd t}
&=&
\hbar\,\omega\,\Gamma
\sum_{\nn}
\big\{
\sum_{i}
2 n_i(n_i-1)\,\rho_{\nn,\nn}
\nppp&&\hspace{-40pt}
+
\sum_{i\neq j}
\big[
3 n_i n_j\,\rho_{\nn,\nn}
-
\sqrt{n_i n_j(n_i-1)(n_j-1)}\,
\nppp&&\hspace{-40pt}
\times\;
\rho_{\nn-2\hat\nn_i,\nn-2\hat\nn_j}
\big]\big\}.
\label{pqdis}
\end{eqnarray}
This expression indeed agrees with the quadrupole radiation formula Eq.~\eqref{qhpwr} applied to the present configuration and quantized with consistent factor ordering described in Eq.~\eqref{anij} where the role of $a_{ij}$ is played by the Lindblad operators $A_{ij}$ here. See \apx{app:agr} for an explicit proof.

By virtue of its inherent Lindblad structure, the master equation~\eqref{maseq} is capable of generating new nonlinear collective quantum gravity phenomena transferred and inspired from more established quantum optics areas sharing similar dynamical structures.

One such novel effect is the collectively amplified spontaneous emission of gravitons by a matter system in a highly coherent state, akin to Dicke's superradiance~\cite{Dicke1954}. To illustrate this, let us consider the present harmonic potential containing many particles excited in one direction, say along the $x$-axis, with a modal occupation state vector denoted by
$\ket{\NN} = \ket{\{N_n\}} = \ket{N_0,N_1,\cdots}$,
where $n=n_1=0,1,2\dots$ labels the harmonic mode in this direction.
It follows that the master equation~\eqref{maseq} has the following matrix elements
\begin{eqnarray}
\bra{\NN}\dot\rho\ket{\NN'}
&=&
\frac{\Gamma}{2}
\Big\{
\DA_{n,n'}(\NN,\NN')
\bra{\NN^{n'+}}\rho\ket{\NN'^{n+}}
\nppp&&\hspace{-45pt}
-
\DB_{n,n'}({\NN})
\bra{\NN^{n-,n'+}}\rho\ket{\NN'}
\Big\}
+(\NN\leftrightarrow\NN')^*
\label{Neq5}
\end{eqnarray}
in terms of nonnegative coefficients
\begin{eqnarray}
\DA_{n,n'}(\NN,\NN')
&=&
\big[N'_{n} N_{n'} (N'_{n+2}+1)(N_{n'+2}+1)
\nppp&&\hspace{-30pt}\times\;
(n+1)(n'+1)(n+2)(n'+2)\big]^{1/2}
\label{AN}
\ppp
\DB_{n,n'}({\NN})
&=&
\big[N_{n+2} N^{n-}_{n'} (N_n+1)(N^{n-}_{n'+2}+1)
\nppp&&\hspace{-30pt}\times\;
(n+1)(n'+1)(n+2)(n'+2)\big]^{1/2}
\label{BN}
\end{eqnarray}
where
$N^{n\pm}_{n'}=
N_{n'}\mp\delta_{n,n'}\pm\delta_{n+2,n'}$.
In this case, even and odd harmonic modes are disjointly coupled within their own parities because of the quadrupole nature of the gravitational waves and symmetry of the potential.

\begin{widetext}

\vspace{-10pt}
\begin{figure}[!ht]
\includegraphics[width=1\linewidth]{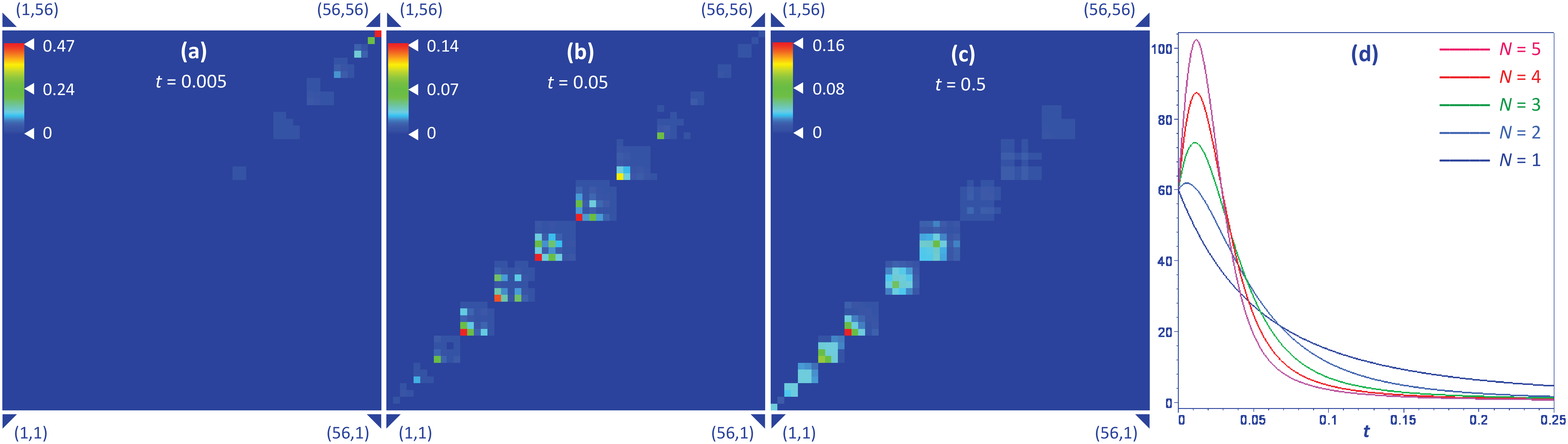}
\caption{Plots (a)--(c) show the simulation of the symmetric modulus of the density matrix  $|\rho_{p,p'}|(t)$  for $p,p'=1,2,\dots 56$ consisting of even harmonic modes for the quantum transitions through the superradiance of gravitational waves. Five scalar bosons in a harmonic trap are initially in the same highest energy state at the top right corner of the density matrix, where all 5 particles occupy the harmonic mode $n=6$. While releasing a short burst of gravitational wave, they spontaneously decay towards the ground state in the bottom left corner, where all 5 particles occupy the $n=0$ mode. Plot~(d) shows the average radiation power per particle as a function of time for a similar initial state as in plots (a)--(c) but with different particle numbers.}
\label{fig1}
\vspace{10pt}
\includegraphics[width=1\linewidth]{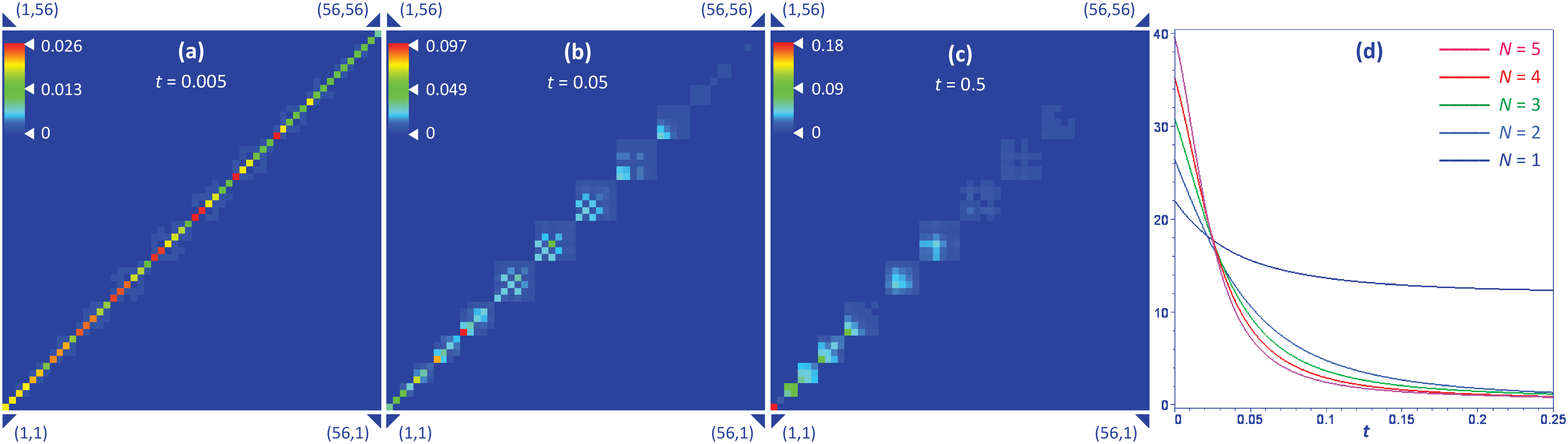}
\caption{Plots (a)--(c) show the simulation of $|\rho_{p,p'}|(t)$. Here 5 scalar bosons in a harmonic trap are initially equally distributed along the diagonal of the density matrix for harmonic modes $n=0,2,4,6$ as a maximally mixed state.  While releasing a continuously decreasing gravitational wave, they spontaneously decay towards the ground state in the bottom left corner, where all 5 particles occupy the $n=0$ mode. Plot~(d) shows the average radiation power per particle for a similar initial state with different particle numbers.}
\label{fig2}
\end{figure}
\vspace{-15pt}
\end{widetext}

Based on the master equation with components~\eqref{Neq5},
we perform numerical simulations in nondimensional time $t \to \Gamma\, t$ initially excited and subsequently relaxed in the $x$-direction, with harmonic modes $n=n_1=0,2,4,6$, $n_2=n_3=0$ and a total particle number $N=1,2,3,4,5$ shown in Figs.~\ref{fig1}--\ref{fig3}. The collective behaviour of the ``superradiant'' spontaneous emission of gravitons, due to the quadratic dependence of the particle occupations (of bosonic origin) as well as modal numbers (of quadrupole origin) in Eqs.~\eqref{AN} and \eqref{BN}, is particularly evident in Fig.~\ref{fig1}, using an initial single-mode Fock state. Milder amplification of emission power with particle numbers are also seen in Figs.~\ref{fig2} and \ref{fig3},  where the initial states may be described as maximally mixed and maximally entangled respectively. The quantum states are enumerated with $\ket{p}$ for $p=1,2,\dots p_{\max}$ with ascending eigen energies and then the particle number occupations of higher harmonic modes. Thus, with $N=5$ there are $p_{\max}=56$ even-mode occupation states $\ket{N_0, N_1\dots,N_6}$ with $N_1=N_3=N_5=0$, starting from the ground state ${\ket{1}=\ket{5,0,0,0,0,0,0}}$, then the first excited state ${\ket{2}=\ket{4,0,1,0,0,0,0}}$ through ${\ket{28}=\ket{2,0,1,0,0,0,2}}$ to the highest state ${\ket{56}=\ket{0,0,0,0,0,0,5}}$.

While classical sources of gravitational waves are of astronomical scales,
the mechanism of collectively enhanced quantum gravitational radiation considered above may open up a future prospect of a lab-sized gravitational wave transmitter. Based on the ongoing rapid development of high-$Q$ nanomechanical resonators demonstrated in the quantum regime\refs{Khalili2010, Chan2011, Safavi2012, Cohen2015}, one could envisage a high-density cluster of nanoresonators in such a correlated state that they behave like a system of $N$ identical harmonic oscillators with frequency $\omega$. Supposing these oscillators occupy around the $n$-th harmonic mode, then following discussions of \Eq{Neq5}, the maximum spontaneous decay rate due to collective gravitational radiation is approximately given by $\Gamma_\text{max} = N^2 n^2 \Gamma$.
For example, a future such microfabricated cluster consisting of up to $N=$ one mole of nanoresonators at $\omega/2\pi=$ 10 GHz excited with $n=1000$ could in principle have an observable peak decay rate of up to $\Gamma_\text{max} =$ 1 Hz via the superradiant spontaneous emissions of gravitons. Furthermore, such gravitons could also be detected using a similar cluster of nanoresonators instead of an ensemble of atoms as a gravitational radiation receiver described in \Ref{Quinones2017}. The quantum nature of gravity could then be probed through the quantum properties of the nanoresonators imparted by the absorbed gravitons.

\begin{widetext}

\vspace{-10pt}
\begin{figure}[!ht]
\includegraphics[width=1\linewidth]{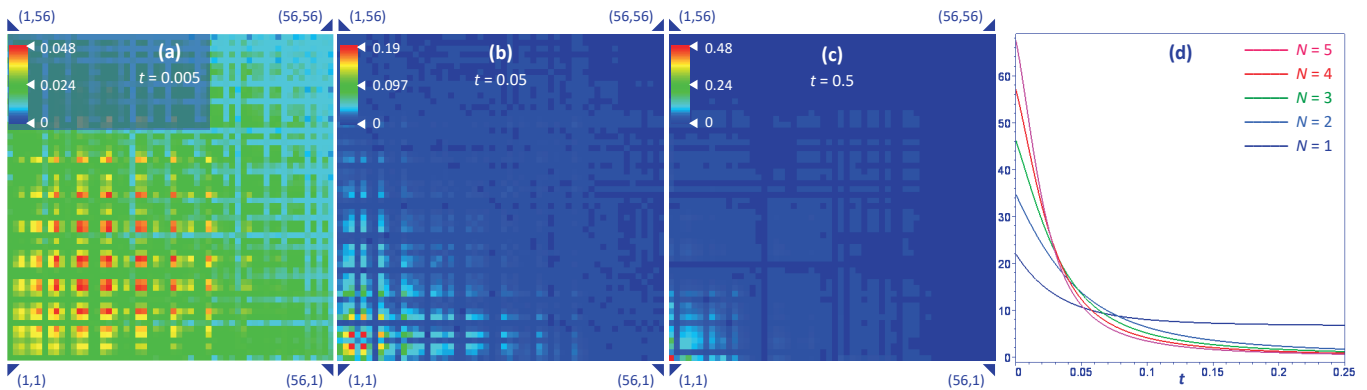}
\caption{
Plots (a)--(c) show the simulation of $|\rho_{p,p'}|(t)$. Here, 5 scalar bosons in a harmonic trap are initially equally and fully distributed in the density matrix for harmonic modes $n=0,2,4,6$ as a maximally entangled state. While releasing a continuously decreasing gravitational wave, they spontaneously decay towards the ground state in the bottom left corner, where all 5 particles occupy the $n=0$ mode. Plot~(d) shows the average radiation power per particle for a similar initial state with different particle numbers.
}
\label{fig3}
\end{figure}
\vspace{-15pt}
\end{widetext}

\section{Conclusion}
\label{sec:con}

Motivated by the need for a better understanding of the fundamental process for quantum matter to decohere and dissipate through spontaneous emission and exchange of gravitons with the ubiquitous fluctuating gravitational environment, we have extended a recently established theory of quantum gravitational decoherence~\cite{Oniga2016a}, now complete with the dynamical origin and consequence of gravitons mediating spacetime at large and matter, both bosons and fermions, of interest.

For physically common states subject to a potential, we have explicitly demonstrated that the abstract master equation describing the general non-Markovian gravitational decoherence of matter formulated in Ref.~\cite{Oniga2016a} can indeed be reduced, free from UV-cutoff, to a more concrete Lindblad form, structurally identical to the family of quantum optical master equations widely applied in the quantum optics problems. This enables investigations of the theory and phenomenology of quantum gravity to benefit from a wealth of novel characteristics and solution strategies in the field of quantum optics~\cite{Breuer2002, Dicke1954, HBT1956, PPT1, PPT2, Metcalf2013}. One such possibility in terms of the newly identified superradiance of gravitational waves by a system of coherence particles has been theoretically described and numerically illustrated in Sec.~\ref{sec:col}.

Our general framework may serve to clarify various conceptual issues encountered in the phenomenological approach to quantum gravity~\cite{Amelino2000, Schiller2004, Lamine2006, Wang2006, Lamine2006, Hu2013, Amelino2013, Ford2015}, with first-principles insights, and to guide further analytical tools,  mathematical techniques, and modelling methodologies for possible detections of quantum gravity effects in the laboratory~\cite{Pfister2016} and observatory~\cite{Vasileiou2015} on the ground or in space~\cite{gauge2009}.
In the context of the cosmological stochastic gravitational waves, since the universe is considered spatially flat with a low entropy on exit from inflation~\cite{Wang2016}, our theory may describe short-time graviton radiation and reception by a distribution of coherent states having potentially unexpected but important collective properties including quantum nonlinearity, nonlocality, and entanglement~\cite{Oniga2016b, Oniga2016c}. In this regard, the theoretical framework reported here has recently been applied and further extended to address the possible detection of stochastic gravitational waves using correlated atoms\refs{Quinones2017} and potential observation of spacetime fluctuations through gravitational lensing\refs{Oniga2017}.

Another future objective would be to go beyond the perturbative formulation so as to accommodate larger spacetime fluctuations and curved background or none.
Extension in this direction could allow the quantum-to-classical transition in the early universe with graviton productions to be more accurately analyzed. This may be initiated by generalizing our non-Markovian master equation\eq{maseqn} to accommodate cosmological perturbations\refs{Sasaki2012}, in addition to its existing gravitational fluctuations in vacuum. A qualitative study of quantum-to-classical transition may follow from the resulting decoherence of the content of the early universe in the presence of cosmological perturbations.
An additional rationale for this final remark is that the development of open quantum gravitational systems towards background independence~\cite{AshtekarLoops1991, Wang2005a, Wang2005b, Wang2006a, Veraguth2017} might even help navigate the search for an ultimate full quantum theory of gravity with compatible and accessible low energy effects like gravitational decoherence and radiance.

\begin{widetext}

\acknowledgments

This work was supported by the Carnegie Trust for the Universities of Scotland (T. O.) and by the EPSRC GG-Top Project and the Cruickshank Trust (C. W.).

\appendix

\renewcommand{\ppp}{\\[1pt]}
\renewcommand{\nppp}{\nonumber\\[1pt]}

\section{Derivation of the gravitational quantum optical master equation}
\label{app:maseq}

To derive \Eq{maseq}, we first introduce
\begin{eqnarray}
\tilde{\tau}_{ij}(\kk,t)
&=&
\int_{0}^{\infty}\dd s\,
\tau_{ij}(\kk,t')
e^{-i k s}
\nppp
&=&
\pi\,\sum_{\nn,\nn'}
F_{ij}(\nn,\nn',\kk)\,
a_{\nn'}^\dag a_{\nn}
e^{-i(\omega_{\nn}-\omega_{\nn'}) t}
\,\delta(k-\omega_{\nn}+\omega_{\nn'})
\label{3ht10f}
\end{eqnarray}
using \Eqq{3ht10d}{thm}.
Note that since $k\ge0$, we have nonzero $\delta(k-\omega_{\nn}+\omega_{\nn'})=0$ only if $\omega_{\nn} \ge \omega_{\nn'}$.
The following relations then hold
\begin{eqnarray}
\tau^\dag_{ij} (\kk,t)
\tilde{\tau}_{ij}(\kk,t)
\,\rho\,
&=&
\sum
\delta(k-\omega(\Delta \nn))\,
A_{ij}^\dag(\nn,\Delta \nn,\kk)\,
A_{ij}(\mm,\Delta \mm,\kk)\,
\rho
\label{3haatt01}
\ppp
\tilde{\tau}_{ij}(\kk,t)
\,\rho\,
\tau^\dag_{ij} (\kk,t)
&=&
\sum
\delta(k-\omega(\Delta \nn))\,
A_{ij}(\mm,\Delta \mm,\kk)
\,\rho\,
A_{ij}^\dag(\nn,\Delta \nn,\kk)
\label{3haatt02}
\ppp
\,\rho\,
\tilde{\tau}_{ij}(\kk,t)
\tau^\dag_{ij} (\kk,t)
&=&
\sum
\delta(k-\omega(\Delta \nn))
\,\rho\,
A_{ij}(\mm,\Delta \mm,\kk)\,
A_{ij}^\dag(\nn,\Delta \nn,\kk)
\label{3haatt03}
\ppp
\tau^\dag_{ij} (\kk,t)
\,\rho\,
\tilde{\tau}_{ij}(\kk,t)
&=&
\sum
\delta(k-\omega(\Delta \nn))\,
A_{ij}^\dag(\nn,\Delta \nn,\kk)
\,\rho\,
A_{ij}(\mm,\Delta \mm,\kk)
\label{3haatt04}
\end{eqnarray}
summing over $\nn,\mm,\Delta \nn,\Delta \mm$ subject to
$\omega(\Delta \nn)=\omega(\Delta \mm)$
in terms of the operators
\begin{eqnarray}
A_{ij}(\nn,\Delta \nn,\kk)
=
\sqrt{{\pi}} F_{ij}(\nn+\Delta \nn,\nn,\kk)\, a_{\nn}^\dag a_{\nn+\Delta \nn}.
\label{3hAop}
\end{eqnarray}
Using Eqs. \eqref{3haatt01}--\eqref{3haatt04}, we have
\begin{eqnarray}
&&\hspace{-20pt}
[
\tau^\dag_{ij} (\kk,t),\,
\tilde\tau_{ij}(\kk,t) \rho
]
=
\tau^\dag_{ij} (\kk,t)
\tilde\tau_{ij}(\kk,t)
\rho
-
\tilde\tau_{ij}(\kk,t)
\rho
\tau^\dag_{ij} (\kk,t)
\nppp
&=&
\sum
\delta(k-\omega(\Delta\nn))
\big[
A_{ij}^\dag(\nn,\Delta \nn,\kk)\,
A_{ij}(\mm,\Delta \mm,\kk)\,
\rho
-
A_{ij}(\mm,\Delta \mm,\kk)
\,\rho\,
A_{ij}^\dag(\nn,\Delta \nn,\kk)
\big]
\label{3haattr1}
\end{eqnarray}
and
\begin{eqnarray}
&&\hspace{-20pt}
[
\tau^\dag_{ij} (\kk,t),\,
[
\tilde\tau_{ij}(\kk,t),\,
\rho
]]
=
\tau^\dag_{ij} (\kk,t)
\tilde\tau_{ij}(\kk,t)\rho
-
\tilde\tau_{ij}(\kk,t)\rho
\tau^\dag_{ij} (\kk,t)
+
\rho\tilde\tau_{ij}(\kk,t)
\tau^\dag_{ij} (\kk,t)
-
\tau^\dag_{ij} (\kk,t)
\rho\tilde\tau_{ij}(\kk,t)
\nppp
&=&
\sum
\delta(k-\omega(\Delta\nn))
\big[
A_{ij}^\dag(\nn,\Delta \nn,\kk)\,
A_{ij}(\mm,\Delta \mm,\kk)\,
\rho
-
A_{ij}(\mm,\Delta \mm,\kk)
\,\rho\,
A_{ij}^\dag(\nn,\Delta \nn,\kk)
\nppp
&&
+
\,\rho\,
A_{ij}(\mm,\Delta \mm,\kk)\,
A_{ij}^\dag(\nn,\Delta \nn,\kk)
-
A_{ij}^\dag(\nn,\Delta \nn,\kk)
\,\rho\,
A_{ij}(\mm,\Delta \mm,\kk)
\big].
\label{3haattr2}
\end{eqnarray}
From \Eqq{3haattr1}{3haattr2}, we see that
\begin{eqnarray}
&&\hspace{-15pt}
\big[
\tau^\dag_{ij} (\kk,t),\,
\tilde{\tau}_{ij}(\kk,t) \rho(t)
\big]
+
N(k)
\big[
\tau^\dag_{ij} (\kk,t),\,
\big[
\tilde{\tau}_{ij}(\kk,t),\,
\rho(t)
\big]\big]
+\hc
\nppp
&=&
-\sum
\delta(k-\omega(\Delta\nn))\,
\Big[
(1+N(k))\big(
2A_{ij}(\mm,\Delta \mm,\kk)
\rho
A_{ij}^\dag(\nn,\Delta \nn,\kk)
-
\{
A_{ij}^\dag(\nn,\Delta \nn,\kk)
A_{ij}(\mm,\Delta \mm,\kk),
\rho
\}
\big)
\nppp&&
+
N(k)\big(
2A_{ij}^\dag(\nn,\Delta \nn,\kk)
\rho
A_{ij}(\mm,\Delta \mm,\kk)
-
\{
A_{ij}(\mm,\Delta \mm,\kk)
A_{ij}^\dag(\nn,\Delta \nn,\kk),
\rho
\}
\big)
\Big].
\label{3hdisp0}
\end{eqnarray}
Substituting \Eq{3hdisp0} into \Eq{maseqn} with negligible $H_\II^\sys$, we have
\begin{eqnarray}
\dot\rho
&=&
\frac{G}{2\pi^2 \hbar}
\sum
\omega(\Delta \nn)\,
\int\! \dd\Omega(\kk(\Delta \nn))
\nppp&&
\Big[
(1+N(\omega(\Delta \nn)))\big(
2A_{ij}(\nn',\Delta \nn',\kk(\Delta \nn))
\rho
A_{ij}^\dag(\nn,\Delta \nn,\kk(\Delta \nn))
-
\{
A_{ij}^\dag(\nn,\Delta \nn,\kk(\Delta \nn))
A_{ij}(\nn',\Delta \nn',\kk(\Delta \nn)),
\rho
\}
\big)
\nppp&&
+
N(\omega(\Delta \nn))\big(
2A_{ij}^\dag(\nn,\Delta \nn,\kk(\Delta \nn))
\rho
A_{ij}(\nn',\Delta \nn',\kk(\Delta \nn))
-
\{
A_{ij}(\nn',\Delta \nn',\kk(\Delta \nn))
A_{ij}^\dag(\nn,\Delta \nn,\kk(\Delta \nn)),
\rho
\}
\big)
\Big]
\nppp
\label{3hmaseqtube}
\end{eqnarray}
where $\kk(\Delta \nn)$ denotes $\kk$ with $k=\omega(\Delta\nn)$.
Furthermore, from \Eq{3hAop}, we have
\begin{eqnarray*}
A_{ij}(\nn',\Delta \nn',\kk(\Delta \nn))
&=&
\sqrt{{\pi}} F_{ij}(\nn'+\Delta \nn',\nn',\kk(\Delta \nn))\,
A(\nn',\Delta \nn')
\nppp
A_{ij}^\dag(\nn,\Delta \nn,\kk(\Delta \nn))
&=&
\sqrt{{\pi}} F_{ij}^*(\nn+\Delta \nn,\nn,\kk(\Delta \nn))\,
A^\dag(\nn,\Delta \nn)
\end{eqnarray*}
in terms of the operator
\begin{eqnarray}
A(\nn,\Delta \nn)
&=&
a_{\nn}^\dag a_{\nn+\Delta \nn}.
\label{3hAa}
\end{eqnarray}

We then substitute the above into \Eq{3hmaseqtube} to isolate the solid angle integral as follows:
\begin{eqnarray}
\dot\rho
&=&
\sum
F(\nn,\nn',\Delta \nn,\Delta \nn')\Big[
(1+N(\omega(\Delta \nn)))\big(
A(\nn',\Delta \nn')
\rho
A^\dag(\nn,\Delta \nn)
-
\frac12\{
A^\dag(\nn,\Delta \nn)
A(\nn',\Delta \nn'),
\rho
\}
\big)
\nppp&&\hspace{30pt}
+
N(\omega(\Delta \nn))\big(
A^\dag(\nn,\Delta \nn)
\rho
A(\nn',\Delta \nn')
-\frac12
\{
A(\nn',\Delta \nn')
A^\dag(\nn,\Delta \nn)),
\rho
\}
\big)
\Big]
\label{3hmaseqtube2}
\end{eqnarray}
where
\begin{eqnarray}
F(\nn,\nn',\Delta \nn,\Delta \nn')
&=&
\frac{G}{\pi\hbar}
\omega(\Delta \nn)\,
\int\! \dd\Omega(\kk(\Delta \nn))
F_{ij}^*(\nn+\Delta \nn,\nn,\kk(\Delta \nn))
F_{ij}(\nn'+\Delta \nn',\nn',\kk(\Delta \nn))
\label{3hGtube0}
\end{eqnarray}
subject to
$\omega(\Delta \nn)=\omega(\Delta \nn')$.
For $\omega(\Delta \nn) > 0$, we obtain from 
\Eq{d3hFun} that
\begin{eqnarray}
\omega(\Delta \nn)
&=&
2\,\omega
\label{2om}
\end{eqnarray}
which in turn requires
\begin{eqnarray}
\Delta\nn
&=&
2\hat\nn_1,\,2\hat\nn_2,\,2\hat\nn_3,\,
\hat\nn_1+\hat\nn_2,\,
\hat\nn_1+\hat\nn_3,\,
\hat\nn_2+\hat\nn_3
\label{dndn2}
\end{eqnarray}
and, furthermore, the expression
\begin{eqnarray}
F_{ij}(\nn+\Delta\nn,\nn,\kk)
=
-{\hbar\,\omega}\,P_{ijkl}(\kk)f_{kl}(\nn+\Delta\nn,\nn)
\label{3hFuna}
\end{eqnarray}
in terms of
\begin{eqnarray*}
f_{11}(\nn+\Delta\nn,\nn)
&=&
\sqrt{(n_1+1)(n_1+2)}\,
\delta_{\Delta n_1,2}
\delta_{\Delta n_2,0}
\delta_{\Delta n_3,0}
\label{3hFun11a}
\ppp
f_{22}(\nn+\Delta\nn,\nn)
&=&
\sqrt{(n_2+1)(n_2+2)}\,
\delta_{\Delta n_1,0}
\delta_{\Delta n_2,2}
\delta_{\Delta n_3,0}
\label{3hFun22a}
\ppp
f_{33}(\nn+\Delta\nn,\nn)
&=&
\sqrt{(n_3+1)(n_3+2)}\,
\delta_{\Delta n_1,0}
\delta_{\Delta n_2,0}
\delta_{\Delta n_3,2}
\label{3hFun33a}
\ppp
f_{12}(\nn+\Delta\nn,\nn)
&=&
\sqrt{(n_1+1)(n_2+1)}\,
\delta_{\Delta n_1,1}
\delta_{\Delta n_2,1}
\delta_{\Delta n_3,0}
\label{3hFun12a}
\ppp
f_{13}(\nn+\Delta\nn,\nn))
&=&
\sqrt{(n_1+1)(n_3+1)}\,
\delta_{\Delta n_1,1}
\delta_{\Delta n_2,0}
\delta_{\Delta n_3,1}
\label{3hFun13a}
\ppp
f_{23}(\nn+\Delta\nn,\nn)
&=&
\sqrt{(n_2+1)(n_3+1)}\,
\delta_{\Delta n_1,0}
\delta_{\Delta n_2,1}
\delta_{\Delta n_3,1}.
\label{3hFun23a}
\end{eqnarray*}
Using the above relations, \Eq{3hGtube0} then becomes
\begin{eqnarray}
F(\nn,\nn',\Delta \nn,\Delta \nn')
&=&
\frac{2\,G \hbar\,\omega^3}{\pi}\,
\int\! \dd\Omega(\kk(\Delta \nn))
P_{ijkl}(\kk(\Delta \nn))
f_{ij}(\nn+\Delta\nn,\nn)
f_{kl}(\nn'+\Delta\nn',\nn').
\label{3hGtube1}
\end{eqnarray}
Through the identities
\begin{eqnarray*}
P_{ijkl}(\kk)
&=&
\frac{1}{2}
\Big[
\delta_{ik}\delta_{jl}
+\delta_{il}\delta_{jk}
-\delta_{ij}\delta_{kl}
\Big]
\nppp&&
+
\frac{1}{2k^2}
\Big[
\delta_{ij} k_k k_l
+ \delta_{kl} k_i k_j
- \delta_{jk} k_i k_l
- \delta_{ik} k_j k_l
- \delta_{il} k_j k_k
- \delta_{jl} k_i k_k
\Big]
+
\frac{k_i k_j k_k k_l}{2k^4}
\end{eqnarray*}
and
\begin{eqnarray*}
\int\! \dd\Omega(\kk)\, k_i k_j
&=&
\frac{4\pi k^{2}}{3}\,\delta_{ij}
\nppp
\int\! \dd\Omega(\kk)\, k_i k_j k_k k_l
&=&
\frac{4\pi k^{4}}{15}\,
\Big[
\delta_{ij}\delta_{kl}
+\delta_{ik}\delta_{jl}
+\delta_{il}\delta_{jk}
\Big]
\end{eqnarray*}
we obtain that
\begin{eqnarray}
\int\! \dd\Omega(\kk) P_{ijkl}(\kk)
&=&
\frac{4\pi}{15}\,
\Big[
3\delta_{ik}\delta_{jl}
+3\delta_{il}\delta_{jk}
-2\delta_{ij}\delta_{kl}
\Big].
\label{iPijkl}
\end{eqnarray}
Substituting \eqref{iPijkl} into \eqref{3hGtube1}, we have
\begin{eqnarray}
F(\nn,\nn',\Delta \nn,\Delta \nn')
&=&
\frac{\Gamma}{4}\,
\sum_{i,j,k,l}
\Big[
3\delta_{ik}\delta_{jl}
+3\delta_{il}\delta_{jk}
-2\delta_{ij}\delta_{kl}
\Big]
f_{ij}(\nn+\Delta\nn,\nn)
f_{kl}(\nn'+\Delta\nn',\nn')
\nppp
&=&
\frac{\Gamma}{2}\,
\sum_{i,j}
\Big[
3\,f_{ij}(\nn+\Delta\nn,\nn)
f_{ij}(\nn'+\Delta\nn',\nn')
-
f_{ii}(\nn+\Delta\nn,\nn)
f_{jj}(\nn'+\Delta\nn',\nn')
\Big]
\label{3hGtube3}
\end{eqnarray}
where
$
\Gamma
$
is given by \Eq{Gamma0}.
From \Eq{3hGtube3} we have
\begin{eqnarray}
F(\nn,\nn',2\hat\nn_i,2\hat\nn_i)
&=&
\Gamma\,
f_{ii}(\nn+2\hat\nn_i,\nn)
f_{ii}(\nn'+2\hat\nn_i,\nn'),\;
(i=1,2,3)
\label{3hGii}
\ppp
F(\nn,\nn',2\hat\nn_i,2\hat\nn_j)
&=&
-\frac{\Gamma}{2}\,
f_{ii}(\nn+2\hat\nn_i,\nn)
f_{jj}(\nn'+2\hat\nn_j,\nn'),\;
(i \neq j)
\label{3hGij}
\ppp
F(\nn,\nn',\hat\nn_i+\hat\nn_j,\hat\nn_i+\hat\nn_j)
&=&
3\,\Gamma\,
f_{ij}(\nn+\hat\nn_i+\hat\nn_j,\nn)
f_{ij}(\nn'+\hat\nn_i+\hat\nn_j,\nn'),\;
(i \neq j)
\label{3hGijij}
\ppp
F(\nn,\nn',\Delta \nn,\Delta \nn')
&=&
0,\;
(\text{for other } \Delta \nn,\Delta \nn').
\label{a23}
\end{eqnarray}
For $N(\omega)=0$, the master equation \eqref{3hmaseqtube2} then becomes
\begin{eqnarray}
\dot\rho
&=&
\sum
F(\nn,\nn',\Delta \nn,\Delta \nn')
\Big[
A(\nn',\Delta \nn')
\rho
A^\dag(\nn,\Delta \nn)
-
\frac12\{
A^\dag(\nn,\Delta \nn)
A(\nn',\Delta \nn'),
\rho
\}
\Big]
\label{3hmaseqt}
\end{eqnarray}
which can be expanded as
\begin{eqnarray*}
\dot\rho
&=& 
\Gamma\,
\sum_{\nn,\nn'}\sum_{i}
f_{i}(\nn+2\hat\nn_i,\nn)
f_{i}(\nn'+2\hat\nn_i,\nn')
\Big[
A(\nn',2\hat\nn_i)
\rho
A^\dag(\nn,2\hat\nn_i)
-
\frac12\{
A^\dag(\nn,2\hat\nn_i)
A(\nn',2\hat\nn_i),
\rho
\}
\Big]
\nppp&&
-\frac{\Gamma}{2}\,
\sum_{\nn,\nn'}\sum_{i\neq j}
f_{i}(\nn+2\hat\nn_i,\nn)
f_{j}(\nn'+2\hat\nn_j,\nn')
\Big[
A(\nn',2\hat\nn_j)
\rho
A^\dag(\nn,2\hat\nn_i)
-
\frac12\{
A^\dag(\nn,2\hat\nn_i)
A(\nn',2\hat\nn_j),
\rho
\}
\Big]
\nppp&&
+
3\,\Gamma\,
\sum_{\nn,\nn'}\sum_{i < j}
f_{ij}(\nn+\hat\nn_i+\hat\nn_j,\nn)
f_{ij}(\nn'+\hat\nn_i+\hat\nn_j,\nn')
\nppp&&
\Big[
A(\nn',\hat\nn_i+\hat\nn_j)
\rho
A^\dag(\nn,\hat\nn_i+\hat\nn_j)
-
\frac12\{
A^\dag(\nn,\hat\nn_i+\hat\nn_j)
A(\nn',\hat\nn_i+\hat\nn_j),
\rho
\}
\Big].
\end{eqnarray*}
Therefore by using the above, \Eqs{3hAa} and \eqref{3hFun11a}--\eqref{a23},
we  arrive at the master equation
\begin{eqnarray}
\dot\rho
&=&
\Gamma\,
\sum_{i}
\Big[
A_{ii}
\rho
A^\dag_{ii}
-
\frac12\{
A^\dag_{ii}
A_{ii},
\rho
\}
\Big]
\nppp&&
-\frac{\Gamma}{2}\,
\sum_{i\neq j}
\Big[
A_{jj}
\rho
A^\dag_{ii}
-
\frac12\{
A^\dag_{ii}
A_{jj},
\rho
\}
\Big]
+
\frac{3\,\Gamma}{2}\,\sum_{i \neq j}
\Big[
A_{ij}
\rho
A^\dag_{ij}
-
\frac12\{
A^\dag_{ij}
A_{ij},
\rho
\}
\Big]
\label{mseq}
\end{eqnarray}
which simplifies to the form\eq{maseq}
in terms of the Lindblad operators
given by \Eq{AAij}.

\section{Consistency between quantum gravitational emission dissipation power and quantum quadrupole radiation formula}
\label{app:agr}

\subsection{Quantum emission power dissipation for one-particle states}

By construction of the Lindblad operators \eqref{AAij}, we have the following relations
\begin{eqnarray}
A_{ii}^\dag A_{ii}\ket{\mm}
&=&
m_i(m_i-1)\ket{\mm}
\label{AAa}
\ppp
A_{ii} A_{ii}^\dag\ket{\mm}
&=&
(m_i+1)(m_i+2)\ket{\mm}
\ppp
A_{ii}^\dag A_{jj}\ket{\mm}
&=&
\sqrt{m_j(m_j-1)(m_i+1)(m_i+2)}\, \ket{\mm+2\hat\nn_i-2\hat\nn_j}
\ppp
A_{ii} A_{jj}^\dag\ket{\mm}
&=&
\sqrt{m_i(m_i-1)(m_j+1)(m_j+2)}\, \ket{\mm-2\hat\nn_i+2\hat\nn_j}
\ppp
A_{ij}^\dag A_{ij}\ket{\mm}
&=&
m_i m_j\ket{\mm}
\ppp
A_{ij} A_{ij}^\dag\ket{\mm}
&=&
(m_i+1) (m_j+1)\ket{\mm}
\label{AAz}
\end{eqnarray}
where $i\neq j$ and no sums are implied.

Applying the master equation\eq{maseq} or equivalently\eq{mseq} to the one-particle density matrix $\rho$,
we have
\begin{eqnarray*}
\bra{\nn'} \dot\rho \ket{\nn}
&=& 
\frac{\Gamma}{2}\,
\sum_{i}
\bra{\nn'}
\Big[
2A_{ii}
\rho
A^\dag_i
-
A^\dag_i A_{ii} \rho
-
\rho A_{ii}^\dag A_{ii}
\Big]
\ket{\nn}
-
\frac{\Gamma}{4}\,
\sum_{i\neq j}
\bra{\nn'}
\Big[
2A_{jj}
\rho
A_{ii}^\dag
-
A_{ii}^\dag A_{jj} \rho
-
\rho A_{ii}^\dag A_{jj}
\Big]
\ket{\nn}
\nppp&&  
+
\frac{3\Gamma}{4}\,
\sum_{i\neq j}
\bra{\nn'}
\Big[
2A_{ij}
\rho
A^\dag_{ij}
-
A^\dag_{ij} A_{ij} \rho
-
\rho A^\dag_{ij} A_{ij}
\Big]
\ket{\nn}.
\end{eqnarray*}
Using \Eqs{AAa}--\eqref{AAz}, the above yields the one-particle master equation
\begin{eqnarray}
\bra{\nn'} \dot\rho \ket{\nn}
&=& 
{\Gamma}\,
\sum_{i}
\sqrt{(n'_i+1)(n'_i+2)(n_i+1)(n_i+2)}\,
\rho_{\nn'+2\nn_i,\nn+2\nn_i}
-\frac{\Gamma}{2}\,
\sum_{i}
\big[
n'_i(n'_i-1)+n_i(n_i-1)
\big]\,
\rho_{\nn',\nn}
\nppp&&
-
\frac{\Gamma}{2}\,
\sum_{i\neq j}
\sqrt{(n'_j+1)(n'_j+2)(n_i+1)(n_i+2)}\,
\rho_{\nn'+2\nn_j,\nn+2\nn_i}
\nppp&&
+
\frac{\Gamma}{4}\,
\sum_{i\neq j}
\sqrt{n'_i(n'_i-1)(n'_j+1)(n'_j+2)}
\,\rho_{\nn'-2\nn_i+2\nn_j,\nn}
+
\frac{\Gamma}{4}\,
\sum_{i\neq j}
\sqrt{n_i(n_i-1)(n_j+1)(n_j+2)}
\,\rho_{\nn',\nn-2\nn_i+2\nn_j}
\nppp&&
+
\frac{3\Gamma}{2}\,
\sum_{i\neq j}
\sqrt{(n'_i+1)(n'_j+1)(n_i+1)(n_j+1)}\,
\rho_{\nn'+\nn_i+\nn_j,\nn+\nn_i+\nn_j}
-
\frac{3\Gamma}{4}\,
\sum_{i\neq j}
\big[
n'_i n'_j + n_i n_j
\big]\,
\rho_{\nn',\nn}.
\label{drhn}
\end{eqnarray}
To obtain the power dissipation for one-particle states, we first use the system Hamiltonian
\begin{eqnarray}
H_S
&=&
\hbar\,\omega
\sum_{i}
\sum_{\nn}
n_i\,a_{\nn}^\dag a_{\nn}
\label{Hsys}
\end{eqnarray}
derived from $T^{00}$ of the scalar field $\phi$ in the nonrelativistic limit
and master equation \eqref{drhn}. This gives
\begin{eqnarray}
\frac{\dd}{\dd t}\av{H_S}
&=&
\sum_{\nn'} \bra{\nn'} (H_S\dot\rho) \ket{\nn'}
\nppp
&=&
{\hbar\,\omega\Gamma}\,
\sum_{i,k}\sum_{\nn}
(n_k-2\delta_{ik})n_i(n_i-1)\,
\rho_{\nn,\nn}
-
\hbar\,\omega\Gamma\,
\sum_{i,k}\sum_{\nn}
n_kn_i(n_i-1)\,
\rho_{\nn,\nn}
\nppp&&
+
\frac{3\hbar\,\omega\Gamma}{2}\,
\sum_{i\neq j,k}\sum_{\nn}
(n_k-\delta_{ik}-\delta_{jk})n_i n_j\,
\rho_{\nn,\nn}
-
\frac{3\hbar\,\omega\Gamma}{2}\,
\sum_{i\neq j,k}\sum_{\nn}
n_k n_i n_j\,
\rho_{\nn,\nn}\nppp&&
-
\frac{\hbar\,\omega\Gamma}{4}\,
\sum_{i\neq j,k}\sum_{\nn}
(n_k-2\delta_{ik}-2\delta_{jk})\sqrt{n_i(n_i-1)(n_j+1)(n_j+2)}
\,\rho_{\nn,\nn-2\nn_i+2\nn_j}
\nppp&&
+
\frac{\hbar\,\omega\Gamma}{4}\,
\sum_{i\neq j,k}\sum_{\nn}
n_k\sqrt{n_i(n_i-1)(n_j+1)(n_j+2)}
\,\rho_{\nn,\nn-2\nn_i+2\nn_j}
\label{pqdis0}
\end{eqnarray}
yielding the quantum dissipation power through spontaneous emission of gravitons
\begin{eqnarray}
P^{\text{(se)}}
&=&
-\frac{\dd}{\dd t}\av{H_S}
\label{pqh}
\end{eqnarray}
given by \Eq{pqdis}.
Note that the last cross term in \Eq{pqdis0} involving $\rho_{\nn-2\nn_i,\nn-2\nn_j}$ represents a quantum correction of the gravitational wave emission process.

\subsection{Quantum quadrupole radiation formula for one-particle states}

In the nonrelativistic limit, using \Eq{3hphi} we have
\begin{eqnarray*}
I_{ij}
&=&
\int\dd^3x\,x^i x^j
T^{00}
\nppp
&=&
\frac{m}{2}
\sum_{\nn,\nn'}
\Big[
a_{\nn'}^\dag
a_{\nn}
e^{-i(\omega_{\nn}-\omega_{\nn'}) t}
+
a_{\nn}^\dag a_{\nn'}
e^{i(\omega_{\nn}-\omega_{\nn'}) t}
\Big]
\int\dd^3x\,
x^i x^j
\psi_{\nn}(\xx)
\psi_{\nn'}(\xx)
\end{eqnarray*}
yielding
\begin{eqnarray*}
\dddot{I}_{ij}
&=&
4 i \hbar\,\omega^2\,
\Big[
A_{ij}\,e^{-2i\omega t}
-
A_{ij}^\dag\,e^{2i\omega t}
\Big].
\end{eqnarray*}
In terms of the traceless part
$
\dddot{\Idash}_{ij}
=
\dddot{I}_{ij}-\frac13\,\delta_{ij}\dddot{I}_{kk}
$
of the above we obtain and the time averaged product
\begin{eqnarray}
\av{\dddot{\Idash}_{ij}\dddot{\Idash}_{ij}}_\text{time av}
&=& 
16 \hbar^2\omega^4\,
\sum_{i\neq j}
\Big[
A_{ij}^\dag A_{ij}
+
A_{ij}A_{ij}^\dag
\Big]
\nppp&&
-\frac{16}{3}
\hbar^2\omega^4\,
\sum_{i\neq j}
\Big[
A_{i}^\dag A_{j}
+
A_{i}A_{j}^\dag
\Big]
+\frac{32}{3}
\hbar^2\omega^4\,
\sum_{i}
\Big[
A_{i}^\dag A_{i}
+
A_{i}A_{i}^\dag
\Big].
\label{qijqij}
\end{eqnarray}
The one-particle matrix elements of \Eq{qijqij} can be evaluated using \Eqs{AAa}--\eqref{AAz} and applying
consistent factor ordering described in Eq.~\eqref{anij} with respect to the Lindblad operators $A_{ij}$ and $A_{ij}^\dag$  to be
\begin{eqnarray}
\bra{\mm'}\dddot{\Idash}_{ij}\dddot{\Idash}_{ij}\ket{\mm}
&=& 
\frac{32}{3}
\hbar^2\omega^4\,
\sum_{i}
\Big[
m_i(m_i-1)
+
(m_i+1)(m_i+2)
\Big]\,
\delta_{\mm',\mm}
\nppp&&
+16 \hbar^2\omega^4\,
\sum_{i\neq j}
\Big[
m_i m_j
+
(m_i+1) (m_j+1)
\Big]\,
\delta_{\mm',\mm}
\nppp&&
-\frac{16}{3}
\hbar^2\omega^4\,
\sum_{i\neq j}
\sqrt{m_j(m_j-1)(m_i+1)(m_i+2)}\,
\delta_{\mm',\mm+2\hat\nn_i-2\hat\nn_j}
\nppp&&
-\frac{16}{3}
\hbar^2\omega^4\,
\sum_{i\neq j}
\sqrt{m_i(m_i-1)(m_j+1)(m_j+2)}\,
\delta_{\mm',\mm-2\hat\nn_i+2\hat\nn_j}.
\label{mqqm}
\end{eqnarray}

Finally, by using \Eqs{qhpwr}, \eqref{qijqij}, \eqref{AAa}--\eqref{AAz} and \eqref{mqqm} we obtain the quantum quadrupole radiation formula
\begin{eqnarray}
P^\text{(qr)}
&=&
\frac{G}{5}\av{\dddot{\Idash}_{ij}\dddot{\Idash}_{ij}}
\nppp
&=&
\frac{32}{15}
G\hbar^2\omega^4\,
\sum_{\nn}\sum_{i}
\Big[
n_i(n_i-1)
+
(n_i+1)(n_i+2)
\Big]\,
\rho_{\nn,\nn}
+
\frac{16}{5} G\hbar^2\omega^4\,
\sum_{\nn}\sum_{i\neq j}
\Big[
n_i n_j
+
(n_i+1) (n_j+1)
\Big]\,
\rho_{\nn,\nn}
\nppp&&
-\frac{32}{15} G\hbar^2\omega^4\,
\sum_{\nn}\sum_{i\neq j}
\sqrt{n_i n_j(n_i-1)(n_j-1)}\,\rho_{\nn-2\hat\nn_i,\nn-2\hat\nn_j}
\label{pqr}
\end{eqnarray}
which, after rearranging terms, is identical to the quantum emission dissipation power $P^\text{(se)}$ given by \Eq{pqh}. Therefore we have established the agreement between \Eqq{qhpwr}{pqdis}.


\end{widetext}


\begin{thebibliography}{99}

\bibitem{LIGO2016}
B. P. Abbott \etal,
(LIGO Scientific and Virgo Collaborations),
Observation of Gravitational Waves from a Binary Black Hole Merger,
Phys. Rev. Lett. {\bf 116}, 061102 (2016).

\bibitem{Schutz1999}
B. F. Schutz,
Gravitational wave astronomy,
Classical Quantum Gravity {\bf 16}, A131 (1999).

\bibitem{Sesana2016}
A. Sesana,
Prospects for Multiband Gravitational-Wave Astronomy after GW150914,
Phys. Rev. Lett. {\bf 116}, 231102 (2016).

\bibitem{gauge2009}
G. Amelino-Camelia \etal,
GAUGE: The GrAnd Unification Gravity Explorer,
Experimental Astronomy {\bf 23}, 549 (2009), and references therein.


\bibitem{AshtekarLoops1991}
A. Ashtekar, C. Rovelli, and L. Smolin,
Gravitons and loops,
Phys. Rev. D {\bf 44}, 1740 (1991).

\bibitem{Wang2005a}
C. H.-T. Wang,
Conformal geometrodynamics: True degrees of freedom in a truly canonical structure,
Phys. Rev. D {\bf 71}, 124026 (2005).

\bibitem{Wang2005b}
C. H.-T. Wang,
Unambiguous spin-gauge formulation of canonical general relativity with conformorphism invariance,
Phys. Rev. D {\bf 72}, 087501 (2005).

\bibitem{Wang2006a}
C. H.-T. Wang,
New ``phase'' of quantum gravity,
Phil. Trans. R. Soc. A {\bf 364}, 3375 (2006).

\bibitem{kiefer2000}
C. Kiefer, D. Polarski, and A. A. Starobinsky,
Entropy of gravitons produced in the early universe,
Phys. Rev. D 62, 043518 (2000).

\bibitem{Joos2003}
E. Joos, H. D. Zeh, C. Kiefer, D. Giulini, J. Kupsch, and I.-O. Stamatescu,
{\it Decoherence and the Appearance of a Classical World in Quantum Theory} (Springer, Berlin, 2003).


\bibitem{Schlosshauer2008}
M. Schlosshauer, {\it Decoherence and the Quantum-to-Classical Transition} (Springer, Berlin, 2008).

\bibitem{kiefer2012}
C. Kiefer,
Emergence of a classical Universe from quantum gravity and cosmology,
Phil. Trans. R. Soc. A  {\bf 370}, 4566 (2012).

\bibitem{Lim2015}
E. A. Lim,
Quantum information of cosmological correlations,
Phys. Rev. D {\bf 91}, 083522 (2015).


\bibitem{LIGO2016b}
B. P. Abbott \etal,
(LIGO Scientific and Virgo Collaborations),
GW150914: Implications for the Stochastic Gravitational-Wave Background from Binary Black Hole,
Phys. Rev. Lett. {\bf 116}, 131102 (2016).

\bibitem{LIGO2009}
LIGO Scientific and Virgo Collaborations,
An upper limit on the stochastic gravitational-wave background of cosmological origin,
Nature (London) {\bf 460}, 990 (2009).


\bibitem{Oniga2016a}
T. Oniga and C. H.-T. Wang,
Quantum gravitational decoherence of light and matter,
Phys. Rev. D {\bf 93}, 044027 (2016).

\bibitem{Oniga2016b}
T. Oniga and C. H.-T. Wang,
Spacetime foam induced collective bundling of intense fields,
Phys. Rev. D {\bf 94}, 061501(R) (2016).

\bibitem{Oniga2016c}
T. Oniga and C. H.-T. Wang,
Quantum dynamics of bound states under spacetime fluctuations,
J. Phys. Conf. Ser. {\bf 845}, 012020 (2017).

\bibitem{Breuer2002}
H.-P. Breuer and F. Petruccione,
{\it The Theory of Open Quantum Systems}
(Oxford University Press, New York, 2002).

\bibitem{Dicke1954}
R. H. Dicke,
Coherence in spontaneous radiation processes,
Phys. Rev. {\bf 93}, 99 (1954).

\bibitem{Burgess2004}
C.P. Burgess,
Quantum gravity in everyday life: General relativity as an effective field theory,
Living Rev. Relativ. {\bf 7}, 5 (2004), and references therein.

\bibitem{Arteaga2004}
D. Arteaga, R. Parentani, and E. Verdaguer,
Propagation in a thermal graviton background,
Phys. Rev. D {\bf 70}, 044019 (2004).

\bibitem{Blencowe2013}
M. P. Blencowe,
Effective Field Theory Approach to Gravitationally Induced Decoherence,
Phys. Rev. Lett. {\bf 111}, 021302 (2013).

\bibitem{Ashtekar2017}
{\it Loop Quantum Gravity: The First 30 Years},
edited by A. Ashtekar and J. Pullin,
(World Scientific, Singapore, 2017).


\bibitem{Isham1992}
C. J. Isham,
Canonical quantum gravity and the problem of time,
arXiv:gr-qc/9210011.

\bibitem{Hu2013}
C. Anastopoulos and B. L. Hu,
A master equation for gravitational decoherence: Probing the textures of spacetime,
Classical Quantum Gravity {\bf 30},  165007 (2013),
and references therein.


\bibitem{Bassi2017}
A. Bassi, A. Gro{\ss}ardt, and H. Ulbricht,
Gravitational decoherence,
Classical Quantum Gravity {\bf 34}, 193002 (2017),
and references therein.


\bibitem{Penrose1996}
R. Penrose,
On gravity's role in quantum state reduction,
Gen. Relativ. Grav. {\bf 28}, 581 (1996)





\bibitem{Tucker1987}
I. M. Benn and R. W. Tucker,
{\it An Introduction to Spinors and Geometry with Applications in Physics}
(Adam Hilger, Bristol, 1987).

\bibitem{Tucker1995}
R. W. Tucker and C. H.-T. Wang,
Black holes with Weyl charge and non-Riemannian waves,
Classical Quantum Gravity {\bf 12}, 2587 (1995).

\bibitem{MTW1973}
C. W. Misner, K. S. Thorne, and J. A. Wheeler, {\it Gravitation}
(Freeman, New York, 1973).

\bibitem{ADM1962}
R. Arnowitt, R. Deser and C. W. Misner, in {\it Gravitation: An Introduction to Current Research}, edited by L. Witten (Wiley, New York, 1962).

\bibitem{Maggiore2008}
M. Maggiore, {\it Gravitational Waves, Vol. 1: Theory and Experiments}
(Oxford University Press, New York, 2008).

\bibitem{Dirac1964}
P. A. M. Dirac,
{\it Lectures on Quantum Mechanics}
(Yeshiva University, New York, 1964).


\bibitem{Antoniadis1997}
I. Antoniadis, P. O. Mazur, and E. Mottola,
Quantum diffeomorphisms and conformal symmetry,
Phys. Rev. D {\bf 55}, 4756 (1997).



\bibitem{Fleming2012}
C. H. Fleming and B. L. Hu,
Non-Markovian dynamics of open quantum systems: Stochastic equations and their perturbative solutions,
Ann. Phys. (N.Y.) {\bf 327}, 1238 (2012), and references therein.


\bibitem{Schafer1980}
G. Sch\"afer and H. Dehnen,
On the gravitational radiation formula,
J. Phys. A {\bf 13}, 2703 (1980), and references therein.

\bibitem{Schafer1981}
G. Sch\"afer,
Gravitational radiation resistance, radiation damping and field fluctuations,
J. Phys. A {\bf 14}, 677 (1981).

\bibitem{Feynman1963}
R. P. Feynman and F. L. Vernon, Jr.,
The theory of a general quantum system interacting with a linear dissipative system,
Ann. Phys. (N.Y.) {\bf 24}, 118 (1963).




\bibitem{Hu1992}
B. L. Hu, J. P. Paz, and Y. Zhang,
Quantum Brownian motion in a general environment; Exact master equation with nonlocal dissipation and colored noise,
Phys. Rev. D {\bf 45}, 2843 (1992)



\bibitem{Ackerhalt1973}
J. R. Ackerhalt, P. L. Knight, and J. H. Eberly,
Radiation Reaction and Radiative Frequency Shifts,
Phys. Rev. Lett. {\bf 30}, 456 (1973).

\bibitem{DDC1982}
J. Dalibard, J. Dupont-Roc, and C. Cohen-Tannoudji,
Vacuum fluctuations and radiation reaction: Identification of tlieir respective contributions,
J. Physique {\bf 43}, 1617 (1982).

\bibitem{DDC1984}
J. Dalibard, J. Dupont-Roc, and C. Cohen-Tannoudji,
Dynamics of a small system coupled to a reservoir: Reservoir fluctuations and self-reaction,
J. Physique {\bf 45}, 637 (1984).

\bibitem{Scully1988}
J. Gea-Banacloche, M. O. Scully, and M. S. Zubairyz,
Vacuum fluctuations and spontaneous emission in quantum optics,
Physica Scripta {\bf T21}, 81 (1988).

\bibitem{Menezes2015}
G. Menezes and N. F. Svaiter,
Vacuum fluctuations and radiation reaction in radiative processes of entangled states,
Phys. Rev. A 92, 062131 (2015), and references therein.

\bibitem{Quinones2017}
D. A. Qui\~nones, T. Oniga, B. T. H. Varcoe, and C. H.-T. Wang,
Quantum principle of sensing gravitational waves: From the zero-point fluctuations to the cosmological stochastic background of spacetime,
Phys. Rev. D {\bf 96}, 044018 (2017).

\bibitem{Khalili2010}
F. Khalili, S. Danilishin, H. Miao, H. M\"uller-Ebhardt, H. Yang, and Y. Chen,
Preparing a Mechanical Oscillator in Non-Gaussian Quantum States,
Phys. Rev. Lett. {\bf 105}, 070403 (2010).

\bibitem{Chan2011}
J. Chan \etal,
Laser cooling of a nanomechanical oscillator into its quantum ground state
Nature (London) {\bf 478}, 89 (2011).

\bibitem{Safavi2012}
Amir H. Safavi-Naein \etal,
Observation of Quantum Motion of a Nanomechanical Resonator
Phys. Rev. Lett. {\bf 108}, 033602 (2012).

\bibitem{Cohen2015}
J. D. Cohen \etal,
Phonon counting and intensity interferometry of a nanomechanical resonator,
Nature (London) {\bf 520}, 522 (2015).

\bibitem{PPT1}
A. Peres,
Separability Criterion for Density Matrices,
Phys. Rev. Lett. {\bf 77}, 1413 (1996).

\bibitem{PPT2}
M. Horodecki, P. Horodecki, and R. Horodecki,
Separability of Mixed states: Necessary and sufficient conditions,
Phys. Lett. A {\bf 223}, 1 (1996).

\bibitem{HBT1956}
R. Hanbury Brown and R. Q. Twiss,
Correlation between photons in two coherent beams of light,
Nature (London) {\bf 177}, 27 (1956).

\bibitem{Metcalf2013}
H. J. Metcalf,
{\it Laser Cooling and Trapping}
(Springer, New York, 2013), and references therein.


\bibitem{Amelino2000}
G. Amelino-Camelia,
Gravity-wave interferometers as probes of a low-energy effective quantum gravity,
Phys. Rev. D {\bf 62}, 024015 (2000).

\bibitem{Schiller2004}
S. Schiller, C. Lammerzahl, H. Muller, C. Braxmaier, S. Herrmann, and A. Peters,
Experimental limits for low-frequency space-time fluctuations from ultrastable optical resonators,
Phys. Rev. D {\bf 69},  027504 (2004).

\bibitem{Wang2006}
C. H.-T. Wang, R. Bingham, and J. T. Mendonca,
Quantum gravitational decoherence of matter waves,
Classical Quantum Gravity {\bf 23},  L59 (2006).

\bibitem{Lamine2006}
B. Lamine, R. Herve, A. Lambrecht, and S. Reynaud,
Ultimate Decoherence Border for Matter-Wave Interferometry,
Phys. Rev. Lett. {\bf 96},  050405 (2006).

\bibitem{Amelino2013}
G. Amelino-Camelia,
Quantum-spacetime phenomenology,
Living Rev. Relativ. {\bf 16}, 5 (2013),
and references therein.

\bibitem{Ford2015}
V. A. De Lorenci and L. H. Ford,
Decoherence induced by long wavelength gravitons,
Phys. Rev. D {\bf 91}, 044038 (2015).


\bibitem{Pfister2016}
C. Pfister \etal,
A universal test for gravitational decoherence,
Nat. Comms. {\bf 7}, 13022 (2016).

\bibitem{Vasileiou2015}
V. Vasileiou, J. Granot, T. Piran, and G. Amelino-Camelia,
A Planck-scale limit on spacetime fuzziness,
Nat. Phys. {\bf 11},  344 (2015).

\bibitem{Wang2016}
C. H.-T. Wang, J. A. Reid, A. St.J. Murphy, D. Rodrigues, M. Al Alawi, R. Bingham, J. T. Mendon\c{c}a, and T. B. Davies,
A consistent scalar-tensor cosmology for inflation, dark energy and the Hubble parameter,
Phys. Lett. A {\bf 380}, 3761 (2016), and references therein.

\bibitem{Oniga2017}
T. Oniga, E. Mansfield, and C. H.-T. Wang,
Cosmic quantum optical probing of quantum gravity through a gravitational lens,
arXiv:1703.01272.

\bibitem{Sasaki2012}
M. Sasaki,
Inflation and Birth of Cosmological Perturbations,
Fundam. Theor. Phys. {\bf 177}, 305 (2014).

\bibitem{Veraguth2017}
O. J. Veraguth and C. H.-T. Wang,
Immirzi parameter without Immirzi ambiguity: Conformal loop quantization of scalar-tensor gravity,
arXiv:1705.09141 [Phys. Rev. D (to be published)].

\end{thebibliography}
\end{document}